\documentclass[12pt]{article}  

\usepackage{epsf}  
\usepackage{amsbsy}  

\newcommand{\beq}{\begin{equation}}  
\newcommand{\eeq}{\end{equation}}  
\newcommand{\beqa}{\begin{eqnarray}}  
\newcommand{\eeqa}{\end{eqnarray}}

\newcommand{\no}{\nonumber}

\newcommand{\vs}{\vspace{-0.25cm}}  
\newcommand{\dfrac}{\displaystyle \frac}  
 
\begin{document}  
     
\hfill {\small FZJ-IKP(TH)-2000-15}

\bigskip\bigskip\bigskip  
  
\begin{center}  

{{\Large\bf Low energy analysis of the nucleon \\[0.3em]  
   electromagnetic form factors\footnote{Work supported in part 
   by funds provided by the Graduiertenkolleg ``Die Erforschung subnuklearer 
   Strukturen der Materie'' at Bonn University.} 
}}  
  
\end{center}  
  
\vspace{.2in}  
  
\begin{center}  
{\large  Bastian Kubis\footnote{email: b.kubis@fz-juelich.de}, 
Ulf-G. Mei{\ss}ner\footnote{email: Ulf-G.Meissner@fz-juelich.de}}

\bigskip  
  
\bigskip  
  
{\it Forschungszentrum J\"ulich,   
Institut f\"ur Kernphysik (Theorie)\\   
D--52425 J\"ulich, Germany}

\end{center}  
  
\vspace{.7in}  
  
\thispagestyle{empty}   
  
\begin{abstract}  
\noindent  
We analyze the electromagnetic  form factors of the nucleon  
to fourth order in relativistic baryon chiral perturbation theory.   
We employ the recently proposed infrared regularization scheme and show that  
the convergence of the chiral expansion is improved as compared to the  
heavy fermion approach. We also discuss the inclusion of vector mesons  
and obtain an accurate description of all four nucleon form factors  
for momentum transfer squared up to $Q^2 \simeq 0.4\,$GeV$^2$.  
\end{abstract}  
  
\vspace{1.3in}  
  
\centerline{PACS: 12.39.Fe, 13.40.Gp, 14.20.Dh}  
  
\centerline{Keywords:   
{\it Nucleon electromagnetic form factors}, {\it chiral perturbation theory}}  
  
\vfill

\newpage  
  
\section{Introduction}  
\label{sec:intro}  
\def\theequation{\arabic{section}.\arabic{equation}}  
\setcounter{equation}{0}  

The electromagnetic structure of the nucleon as revealed in elastic
electron--nucleon scattering is parameterized in terms of  four
form factors.\footnote{As will be discussed in more detail later,
one can either work with the Dirac and Pauli form factors $F_1$ and $F_2$
or the Sachs form factors $G_E$ and $G_M$ of the proton and the neutron.}
The understanding of these form factors is of utmost
importance in any theory or model of the strong interactions.
Abundant data on these form factors over a large range of momentum
transfer already exist, and this data base will considerably improve in
the few GeV region as soon as further experiments at CEBAF will be completed and
analyzed. In addition, experiments involving polarized beams and/or
targets are also performed at lower energies to give better data in
particular for the electric form factor of the neutron, but also for the
magnetic proton and neutron ones. Such kinds of experiments have been performed
or  are under way at NIKHEF, MAMI, ELSA, MIT--Bates and other places. 
Clearly, theory has to provide
a tool to interpret these data in a model--independent fashion.
For small momentum transfer, this can be done in the framework of baryon
chiral perturbation theory (ChPT), which is the effective field theory of the
Standard Model at low energies. This will be the main topic of the present
investigation.

\medskip\noindent
To put our work presented here into a  better perspective, let us recall what
is already known from chiral perturbation theory studies of the 
electromagnetic form factors.
Already a long time ago it was established that the isovector (Dirac and
Pauli) charge radii diverge in the chiral limit of vanishing pion
mass~\cite{BZ}. The first systematic investigation in the framework of
relativistic baryon ChPT was given in~\cite{GSS}, in particular, it was shown
that the analytic structure of the one--loop representation of the isovector
spectral functions is in agreement with the one deduced from unitarity
(in the low energy region, that is on the left wing of the rho resonance). That
approach, however, suffers from the fact that due to the use of standard dimensional
regularization, the one--to--one correspondence between the expansion in loops
and the one in small momenta was upset. If one considers the nucleon as a very
heavy static source, a consistent power counting is possible, the so--called
heavy baryon chiral perturbation theory (HBChPT). Within that approach, the
electromagnetic form factors were studied in \cite{BKKM,BFHM}, the latter
also containing the extension to an effective field theory including the delta
resonance. It was found that the chiral description already fails for values of the
four--momentum transfer squared of about  $Q^2 \simeq 0.2\,$GeV$^2$.
In \cite{spectral}, the isovector and isoscalar spectral functions
were investigated in the heavy nucleon approach. It was pointed out that due
to the heavy mass expansion, the analytic structure of the isovector spectral
function is distorted, making the chiral expansion fail to converge in certain
regions of small momentum transfer.  This can be overcome in the recently
proposed Lorentz--invariant formulation of~\cite{Becher} making use of the
so--called ``infrared regularization''. Being relativistic, 
this approach leads by construction to
the correct behavior of the spectral functions in the low energy
domain. Furthermore, it is expected to improve the convergence of the chiral
expansion. This will be one of the main issues to be addressed here.
We perform a complete one--loop analysis of the form factors, i.e.\ taking into
account all terms up to fourth order.
We will demonstrate that one can achieve a good description of the neutron
charge form factor for momentum transfer squared up to about $Q^2 =
0.4\,$GeV$^2$. The other three form factors cannot be precisely reproduced
by the pion cloud plus local contact terms to this order. The source of this
deficiency is readily located -- it stems from the contribution of vector
mesons. We show how to incorporate these in a chirally symmetric manner
without introducing additional free parameters (the new parameters related
to the vector mesons are taken from dispersion--theoretical analyses of the
form factors, see~\cite{hoehler,mmd,hmd}). Within that framework, we obtain
a very precise description of the large ``dipole--like'' form factors without
destroying the good result for the neutron form factor obtained in the pure
chiral expansion. It is also important to stress that the results obtained
in the heavy baryon approach can be straightforwardly deduced from the
framework employed here, shedding some more light on previous HBChPT results.

\medskip\noindent  
The manuscript is organized as follows. 
In section~\ref{sec:ff}, some basic definitions concerning the electromagnetic 
form factors are collected.
The one--loop representation of the nucleon form factors is given in section~\ref{sec:1loop}.  
The effective Lagrangian on which our investigation is based is briefly reviewed 
in subsection~\ref{sec:lagr}. 
In particular, infrared regularization and the treatment of loop integrals are 
discussed in some detail in subsection~\ref{sec:IR}. 
The pertinent results based on this complete one--loop representation are presented 
and discussed in subsections~\ref{sec:chiexpff} and \ref{sec:res}. 
Particular emphasis is put on a direct comparison with the results obtained in heavy baryon 
chiral perturbation theory, which is a limiting case of the procedure employed in this 
manuscript.  
The inclusion of vector mesons in harmony with chiral symmetry is presented  
in section~\ref{sec:vectors}. 
We give some technicalities in subsection~\ref{sec:tensor} 
and display the results for the form factors in \ref{sec:Vres}, 
followed by some remarks on resonance saturation in \ref{sec:ressat}. 
Concluding remarks are given in section~\ref{sec:sum}.  
Some further technical aspects are relegated to the appendices.

\section{Nucleon form factors}  
\label{sec:ff}  
\def\theequation{\arabic{section}.\arabic{equation}}  
\setcounter{equation}{0}

The structure of the nucleon (denoted by `$N$') as probed by virtual photons  
is parameterized in terms of four form factors,  
\beq  
\langle N(p')\, | \, {\cal J}_\mu \,  | \, N(p)\rangle   
= e \,  \bar{u}(p') \, \biggl\{  \gamma_\mu F_1^{N} (t)  
+ \frac{i \sigma_{\mu \nu} q^\nu}{2 m_N} F_2^{N} (t) \biggr\}   
\,  u(p) \,, \quad N=p,n \,,  
\eeq  
with $t = q_\mu q^\mu = (p'-p)^2$ the invariant momentum   
transfer squared, ${\cal J}_\mu$   
the isovector vector quark current,
${\cal J}_\mu = \bar{q} {\cal Q} \gamma_\mu q$ (${\cal Q}$ is the quark charge matrix),
and $m_N$ the mean nucleon mass.  
In electron scattering, $t$ is negative and it is often convenient   
to define the positive  
quantity $Q^2 = -t > 0$. $F_1$ and $F_2$ are called the Dirac and the Pauli  
form factor, respectively, with the normalizations $F_1^p (0) =1$,  
$F_1^n (0) =0$, $F_2^p (0) =\kappa_p$ and $F_2^n (0) =\kappa_n$. Here,  
$\kappa$ denotes the anomalous magnetic moment.  
One also uses the electric and magnetic Sachs form factors,  
\beq  
G_E (t) = F_1 (t) + \dfrac{t}{4m_N^2} F_2 (t) \, , \quad G_M (t) = F_1 (t) + F_2 (t) \, .  
\label{sachsdef}
\eeq  
In the Breit--frame, $G_E$ and $G_M$ are nothing but the Fourier--transforms  
of the charge and the magnetization distribution, respectively.  
In a relativistic framework, as it is used throughout this text, the Dirac  
and Pauli form factors arise if one constructs the most  
general nucleonic matrix element of the electromagnetic current consistent  
with Lorentz invariance, parity and charge conjugation. In the  
non--relativistic limit on the other hand, in which the nucleon can be considered as a very  
heavy static source, one naturally deals with the Sachs form  
factors. Therefore, as first stressed in~\cite{BKKM}, these arise in the  
heavy baryon approach to chiral perturbation theory. Since we will frequently  
compare our results to those obtained in that framework, we will  consider both  
Pauli and Dirac and the Sachs form factors.  
For the theoretical analysis, it is advantageous to work in the isospin basis,   
\beq  
F_i^{s,v} (t) = F_i^p (t) \pm F_i^n (t)~, \quad (i=1,2)~,   
\eeq  
since the photon has an isoscalar ($I=s$) and an isovector ($I=v$)   
component (and similarly for the Sachs form factors).  
It is important to note that the lowest hadronic states to which the  
isoscalar and isovector photons couple are the two and three pion systems,  
respectively. These are the corresponding thresholds for the absorptive  
parts of the isoscalar and isovector nucleon form factors.  
  
\medskip\noindent  
The slope of the form factors at $t=0$ is conventionally expressed in  
terms of a nucleon radius $\langle r^2\rangle^{1/2}$,  
\beq  
F(t) = F(0) \, \biggl( 1 + \dfrac{1}{6}\langle r^2\rangle \, t + \ldots \biggr) \,  
\eeq  
which is rooted in the non--relativistic description of the scattering  
process in which a point--like charged particle interacts with a  
given charge distribution $\rho(r)$. The mean square radius of this  
charge distribution is given by  
\beq  
\langle r^2\rangle = 4 \pi\, \int_0^\infty dr \,  r^2 \rho(r) = -  
\dfrac{6}{F(0)} \dfrac{dF(Q^2)}{dQ^2} \biggr|_{Q^2 = 0} \, \, .  
\label{defrad}    
\eeq  
Eq.~(\ref{defrad}) can be used for all form factors except $G_E^n$ and  
$F_1^n$ which vanish at $t=0$. In these cases, one simply drops the  
normalization factor $1/F(0)$ and defines e.g.\ the neutron charge radius  
via  
\beq  
\langle (r_E^n)^2 \rangle \,  = -6 \, \dfrac{dG_E^n(Q^2)}{dQ^2}  
\biggr|_{Q^2 = 0} \, \, . \eeq  
It is important to note that the slopes of $G_E^n$ and  
$F_1^n$ are related via  
\beq  
\dfrac{dG_E^n(Q^2)}{dQ^2} \biggr|_{Q^2 = 0} = \dfrac{dF_1^n(Q^2)}{dQ^2}  
\biggr|_{Q^2 = 0} - \dfrac{F_2^n (0)}{4m_N^2} \, \, ,  
\label{foldy}     
\eeq  
where the second term in eq.~(\ref{foldy}) is called the Foldy term. It gives  
the dominant contribution to the slope of $G_E^n$.

\medskip\noindent  
The large body of electron scattering data spanning momentum transfers from  
$Q^2 \simeq 0$ to $Q^2 \simeq 35\,$GeV$^2$ can be analyzed in a largely  
model--independent fashion using dispersion relations~\cite{hoehler,mmd,hmd}.  
Here, we are interested in the region of small momentum transfer, $Q^2 \le  
0.5\,$GeV$^2$. Nevertheless, since there is still some substantial scatter in  
the data in this range of momentum transfer, we will also use the results of  
the dispersive analysis for comparison to the ones obtained in the chiral expansion.  
A recent review on the theory of the form factors is given  
in~\cite{ulfn99}, the status of the data as of 1999 is discussed in~\cite{petn99}.   
  
\section{One--loop representation}
\label{sec:1loop}
\def\theequation{\arabic{section}.\arabic{equation}}  
\setcounter{equation}{0}  


\subsection{Effective Lagrangian}  
\label{sec:lagr}  
  
Our starting point is an effective field theory of asymptotically observable  
relativistic spin--1/2 fields, the nucleons, chirally coupled to the   
pseudo--Goldstone bosons of QCD, the pions, and external sources (like e.g.\ the  
photon field).  The theory shares the  
symmetries of QCD (spontaneously and explicitly broken chiral symmetry,  
parity, charge conjugation, time reversal invariance, and Lorentz covariance) and  
can be formulated in such a way as to obey  systematic power counting, 
which will be discussed at length in the following section.
External momenta and quark (pion) mass insertions  
are treated as small quantities in comparison to the scale of chiral symmetry  
breaking, $\Lambda_\chi \simeq 1\,$GeV. The most direct link of the effective  
field theory to the underlying one, QCD, comes from the analysis of the chiral  
Ward identities as stressed by Leutwyler~\cite{heiri}. Any contribution to 
S--matrix elements or transition currents has the form  
\beq  
{\cal M} = q^\nu \, f\left( {q \over \lambda}, g \right)~,  
\eeq  
where $q$ is a generic symbol for any small quantity, $f$ a function of order  
one, $\lambda$ a regularization scale, and $g$ a collection of coupling constants.  
Chiral symmetry   demands that the power $\nu$ is bounded from below.  The  
expansion in increasing powers of this parameter is called the {\it chiral}  
expansion. At a given order, one has to consider tree as well as loop  
diagrams, the latter ones restoring unitarity in a perturbative fashion.  
The machinery to perform this is based on an effective Lagrangian,  
which consists of a string of terms of increasing (chiral) dimension,  
\beq\label{Lagr}  
{\cal L}_{\rm eff} = {\cal L}_{\pi\pi}^{(2)} +
 {\cal L}_{\pi N}^{(1)} +  {\cal L}_{\pi N}^{(2)} +  
 {\cal L}_{\pi N}^{(3)} +  {\cal L}_{\pi N}^{(4)} + \ldots~,  
\eeq  
where the ellipsis denotes terms of higher order not needed here. We note that  
we perform a complete one--loop analysis, i.e.\ taking tree graphs with  
insertions from all terms indicated in eq.~(\ref{Lagr}) and {\it one}--loop  
diagrams with insertions from ${\cal L}_{\pi N}^{(1)}$ and at most one  
insertion from ${\cal L}_{\pi N}^{(2)}$.  

\medskip\noindent  
We will now consider the terms relevant to our problem 
(for a more detailed description, we refer the reader to~\cite{BKMrev}). 
The chiral effective pion Lagrangian, which to leading order  
contains two parameters, the pion decay constant (in the chiral limit) $F$
and the pion mass (its leading term in the quark mass expansion) $M$, is given by
\beq
{\cal L}_{\pi\pi}^{(2)} = \frac{F^2}{4}\langle u_\mu u^\mu +\chi_+ \rangle ~,
\eeq
where the triplet of pion fields is collected in the SU(2) valued matrix  
$U(x) = u^2(x)$, and the chiral vielbein is related to $u$ via
$u_\mu = i\{u^\dagger , \nabla_\mu u\}$. 
$\nabla_\mu$ is the covariant derivative on the pion fields including
external vector ($v_\mu$) and axial ($a_\mu$) sources,
$\nabla_\mu U= \partial_\mu U -i\bigl(v_\mu+a_\mu\bigr)U+iU\bigl(v_\mu-a_\mu\bigr)$.  
The mass term is included in the field $\chi_+$ via the definitions $\chi = 2B(s+i\,p)$
and $\chi_\pm=u^\dagger \chi u^\dagger \pm u \chi^\dagger u$, with $s$ and $p$ being
scalar and pseudoscalar sources, respectively, the former including the quark mass matrix,
$s={\cal M}+\ldots$~. Furthermore,  $\langle \ldots \rangle$ denotes the trace in flavor space.

\medskip \noindent
The pion--nucleon Lagrangian at leading order reads  
\beq  
{\cal L}_{\pi N}^{(1)} = \bar{\Psi} \Bigl(  
i D\!\!\!\!/ - m + \frac{g_A}{2} u\!\!\!/ \gamma_5 \Bigr) \Psi ~,  
\eeq  
where the bi--spinor $\Psi$ collects the proton and neutron fields and $g_A$  
is the axial--vector coupling constant measured in neutron $\beta$--decay,  
$g_A = 1.26$. To be more precise, the nucleon mass and the axial coupling  
should be taken at their values in the two flavor chiral limit ($m_u = m_d  
=0$, $m_s$ fixed at its physical value). 
To this order, the photon field only couples to the charge of the nucleon.
It resides in the chiral covariant derivative, 
$D_\mu \Psi= \partial_\mu \Psi +\Gamma_\mu \Psi$, 
with the chiral connection given by 
$\Gamma_\mu = \frac{1}{2}\bigl[u^\dagger, \partial_\mu u\bigr]
-\frac{i}{2}u^\dagger (v_\mu+a_\mu)u -\frac{i}{2}u(v_\mu-a_\mu)u^\dagger$.

\medskip \noindent  
The minimal Lagrangians at second and third order have been given in  
\cite{FMS}. We only show the terms needed for the calculation of the form factors,  
\beqa  
{\cal L}_{\pi N}^{(2)} &=& \bar{\Psi} \, \biggl\{  
c_1 \langle \chi_+ \rangle  
-\frac{c_2}{8m^2} \Bigl( \langle u_\mu u_\nu \rangle \{ D^\mu,D^\nu \} +{\rm h.c.} \Bigr)  
+\frac{i\,c_4}{4} \sigma^{\mu\nu} [u_\mu,u_\nu] \no\\  
&& \qquad  
+\frac{c_6}{8m} \sigma^{\mu\nu} F^+_{\mu\nu}   
+\frac{c_7}{8m} \sigma^{\mu\nu} \langle F^+_{\mu\nu} \rangle   
\biggr\} \, \Psi ~, \\  
{\cal L}_{\pi N}^{(3)} &=& \bar{\Psi} \, \biggl\{  
\frac{i\,d_6}{2m}\Bigl( [D^\mu,\hat{F}^+_{\mu\nu}]D_\nu +{\rm h.c.} \Bigr)  
+\frac{i\,d_7}{2m}\Bigl( [D^\mu,\langle F^+_{\mu\nu}\rangle]D_\nu +{\rm h.c.} \Bigr)  
\biggr\} \, \Psi ~.  
\eeqa  
Here, $F^+_{\mu\nu}=u^\dagger F_{\mu\nu} u + u F_{\mu\nu} u^\dagger$, 
and $F_{\mu\nu} = \partial_\mu A_\nu - \partial_\nu A_\mu$
is the conventional photon field strength tensor. Furthermore, we have adopted 
the notation introduced in \cite{FMS} for traceless operators in SU(2), 
$\hat{A}=A-\frac{1}{2}\langle A \rangle$.
The $c_i$ and $d_i$ are the so--called low energy constants (LECs),  
which encode information about the more massive states not contained in the  
effective field theory or other short distance effects. 
These parameters have to be pinned down from some data. 
In our case, the LECs $c_1,\,c_2,\,c_4$, which only appear in loop diagrams, can be 
taken from analyses of $\pi N$--scattering, whereas $c_6,\,c_7$
parameterize the leading magnetic photon coupling to the nucleon, and $d_6,\,d_7$ 
have to be fitted to the charge radii of proton and neutron.

\medskip \noindent  
The minimal pion--nucleon Lagrangian to fourth order has been worked out in \cite{FMMS}.  
Of the 118 terms given there, we only show the four which are of interest here,  
\beqa  
{\cal L}_{\pi N}^{(4)} &=& \bar{\Psi} \, \biggl\{  
-\frac{e_{54}}{2} \bigl[ D^\lambda,[D_\lambda, \langle F^+_{\mu\nu} \rangle ]\,]\,\sigma^{\mu\nu}  
-\frac{e_{74}}{2} \bigl[ D^\lambda,[D_\lambda, \hat{F}^+_{\mu\nu} ]\,]\,\sigma^{\mu\nu} \no\\  
&& \qquad  
-\frac{e_{105}}{2}\langle F^+_{\mu\nu} \rangle \, \langle \chi_+ \rangle \, \sigma^{\mu\nu}  
-\frac{e_{106}}{2}\hat{F}^+_{\mu\nu} \langle \chi_+ \rangle \, \sigma^{\mu\nu}  
\biggr\} \, \Psi ~.  
\eeqa  
Two more terms (numbered 107 and 108 respectively in \cite{FMMS}) only contribute when   
taking into account effects due to $m_u \neq m_d$. We shall disregard these in what follows.  
We also note that the terms $\sim e_{105}, e_{106}$ only amount to a quark mass  
renormalization of the leading magnetic couplings $\sim c_{6},\,c_{7}$, 
\beqa 
c_6 &\to& \tilde{c}_6 = c_6 - 16\, m\, M^2 \, e_{106}~, \no\\
c_7 &\to& \tilde{c}_7 = c_7 - 8\, m\, M^2  \Bigl(2 e_{105} -e_{106} \Bigr)~. 
\label{ctilde}
\eeqa
When it comes to numerical evaluation, we will just use the renormalized constants 
$\tilde{c}_6$ and $\tilde{c}_7$ and will not regard $e_{105}$ and $e_{106}$ as additional
parameters to be fitted.

 
\subsection{Infrared regularization}  
\label{sec:IR} 

\begin{figure}[htb] 
\centerline{ 
\epsfysize=2.6in 
\epsffile{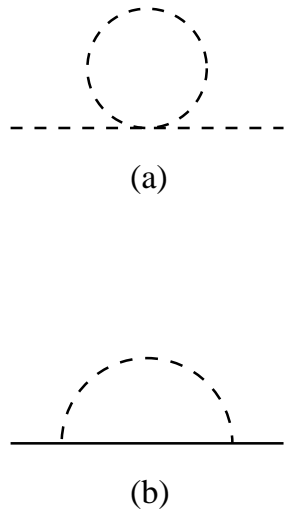} 
}
\vskip 1.3cm 
\caption{Lowest order loop diagrams contributing to the mass renormalization
of (a) the pion at ${\cal O}(q^4)$, and (b) the nucleon at ${\cal O}(q^3)$.
Solid and dashed lines refer to nucleons and pions, respectively.
\label{fig:mass} 
} 
\end{figure}

Chiral perturbation theory in the meson sector as an effective theory for
weakly interacting Goldstone bosons allows for a systematic expansion
of physical observables simultaneously in powers of small momenta and quark masses. 
For example, in the isospin limit ($m_u=m_d$) the quark mass expansion of the
pion mass takes the form
\beq
M_\pi^2 = M^2 \biggl\{ 1 - \frac{M^2}{32\pi^2 F^2} \bar{\ell}_3
\biggr\} + {\cal O}(M^6)~,
\eeq
where the renormalized coupling $\bar{\ell}_3$ depends logarithmically
on the quark mass, $M^2 d\bar{\ell}_3/dM^2$ $= -1$. The infinite contribution
of the pion tadpole $\sim 1/(d-4)$, see graph (a) in fig.~\ref{fig:mass},
has been absorbed in the infinite part of this LEC. In this scheme,
there is a consistent power counting since the only mass scale
(i.e.\ the pion mass) vanishes in the chiral limit $m_u=m_d=0$.
If one uses the same method in the
nucleon case, one encounters a scale problem: for instance,
if the nucleon field is treated relativistically based on standard
dimensional regularization,
the nucleon mass shift calculated from the self--energy diagram
(see graph (b) in fig.~\ref{fig:mass})  can be expressed via \cite{GSS} 
\beq
m_N-m = \frac{3g_A^2 m^2}{32\pi^2 F^2} \, m \,
\biggl\{ \bar{c}_0 + \bar{c}_1 \mu^2 - \pi \mu^3 -  \mu^4 \log \mu
+ \sum_{\nu = 4}^\infty a_\nu \mu^\nu \biggr\}~,
\label{massrel}
\eeq
with $\mu = M/m$, and $\bar{c}_0,\,\bar{c}_1$ are renormalized LECs
(the precise relation of which to previously defined LECs is of no relevance here). 
This expansion is very different from the one for the pion mass, 
in that the nucleon mass already receives a (infinite) renormalization 
in the chiral limit.
This difference is due to the fact that the nucleon mass does not vanish in
the chiral limit and thus introduces a new mass scale apart from the
one set by the quark masses. Therefore, any power of the quark masses
can be generated by chiral loops in the nucleon case, whereas in the
meson case a loop order corresponds to a definite number of quark mass
insertions. This is the reason why one has introduced  the heavy mass
expansion in the nucleon case. Since in that formalism the nucleon mass is
transformed from the propagator into a string of vertices with
increasing powers of $1/m$, a consistent power counting can be formulated.
However, this method has the disadvantage
that certain types of diagrams are at odds with strictures from analyticity.
The best example is the so--called triangle graph, which enters e.g.\ the
scalar form factor or the isovector electromagnetic form factors of the
nucleon (see graph (6) in fig.~\ref{fig:diag}). 
This diagram has its threshold at $t_0= 4M_\pi^2$ but also
a singularity on the second Riemann sheet, at $t_c =  4M_\pi^2 - M_\pi^4/
m_N^2 = 3.98M_\pi^2$, i.e.\ very close to threshold. To leading order
in the heavy baryon approach, this singularity coalesces with the
threshold and thus causes problems (a more detailed discussion can
be found e.g.\ in \cite{spectral,Ulfhugs}).
In a fully relativistic treatment,
such constraints from analyticity are automatically fulfilled. 

\medskip \noindent
It was recently
argued in \cite{EllTa} that relativistic one--loop integrals can be
separated into ``soft'' and ``hard'' parts. While for the former, a similar power counting
as in HBChPT applies, the contributions from the latter can be absorbed in
certain  LECs.
In this way, one can combine the advantages of both methods. A more formal
and rigorous implementation of such a program is due to Becher and
Leutwyler \cite{Becher}. They call their method, 
which we will also use here, ``infrared regularization''. Any
one--loop integral $H$ is split into an infrared singular and a regular part
by a particular choice of Feynman parameterization. Consider first the
regular part, called $R$. If one chirally  expands these terms, one generates
polynomials in momenta and quark masses. Consequently, to any order, $R$ can
be absorbed in the LECs of the effective Lagrangian.  On the other hand, the
infrared singular part $I$ has the same analytical properties as the full
integral $H$ in the low--energy region and its chiral expansion leads to the
non--trivial momentum and quark--mass dependences of ChPT, like e.g.\ the
chiral logarithms or fractional powers of the quark masses.
It is this infrared singular part $I$ that is closely related to the heavy
baryon expansion: there, the relativistic nucleon propagator is replaced
by a heavy baryon propagator plus a series of $1/m$--suppressed 
insertions. Summing up all heavy baryon diagrams with all internal--line insertions
yields the infrared singular part of the corresponding relativistic diagram. 

\medskip \noindent
To be specific, consider the self--energy diagram (b) of fig.~\ref{fig:mass}. 
In $d$ dimensions, the corresponding scalar loop integral is
\beq
H(p^2) = \frac{1}{i} \int \frac{d^dk}{(2\pi)^d} \frac{1}{[M^2 -k^2-i\epsilon]
[m^2 -(p-k)^2-i\epsilon]}~.
\eeq
At threshold, $p^2 = s_0 = (M+m)^2$, this results in
\beq
H(s_0) = c(d)\frac{M^{d-3} + m^{d-3}}{M + m} = I+R~,
\eeq
with $c(d)$ some constant depending on the dimensionality of space--time.
The infrared singular piece $I$ is characterized by fractional powers in the
pion mass and generated by loop momenta of order $M_\pi$. For these soft
contributions, the power counting is fine. On the other hand, the infrared regular
part $R$ is characterized by integer powers in the pion mass and generated
by internal momenta of the order of the nucleon mass (the large mass scale).
These are the terms which lead to the violation of the power counting in
the standard dimensional regularization discussed above. For the self--energy
integral, this splitting can be achieved in the following way:
\beqa
H &=& \int  \frac{d^dk}{(2\pi)^d} {1 \over AB}
= \int_0^1 dz \int  \frac{d^dk}{(2\pi)^d} {1 \over [(1-z)A+zB]^2}
\nonumber \\
 &=& \biggl\{ \int_0^\infty - \int_1^\infty \biggr\} dz
 \int  \frac{d^dk}{(2\pi)^d} {1 \over [(1-z)A+zB]^2} = I + R~,
\eeqa
with $A=M^2-k^2-i\epsilon$, $B=m^2 -(p-k)^2 -i\epsilon$. Any general
one--loop diagram with arbitrary many insertions from external sources
can be brought into this form by combining the propagators to a single
pion-- and a single nucleon--propagator. It was also shown
that this procedure leads to a unique, i.e.\ 
process--independent result, in accordance with the chiral Ward
identities of QCD.  This is essentially based on the fact that terms
with fractional versus integer powers in the pion mass must be separately
chirally symmetric. Consequently, the transition from any one--loop graph $H$
to its infrared singular piece $I$ defines a symmetry--preserving regularization.
However, at present it is not known how to generalize this method to higher
loop orders.  Also, its phenomenological
consequences  have not been explored in great detail so far. It is,
however, expected that this approach will be applicable in a larger energy
range than the heavy baryon approach.
  
\medskip \noindent 
Some remarks concerning renormalization within this scheme are in order. To
leading order, the infrared singular parts coincide with the heavy baryon
expansion, in particular the infinite parts of loop integrals are the same. 
Therefore, the $\beta$--functions for low energy constants which absorb
these infinities are identical. However, infrared singular parts of relativistic
loop integrals also contain infinite parts which are suppressed by powers
of $\mu$, which hence cannot be absorbed as long as one only introduces counter
terms to a finite order: exact renormalization only works up to the order at
which one works, higher order divergences have to be removed by hand.
Closely related to this problem is the one of the new mass scale $\lambda$ which 
one has to introduce in the process of regularization and renormalization.
In dimensional regularization and related schemes, 
loop diagrams depend logarithmically on $\lambda$.
This $\log \lambda$ dependence is compensated for by running coupling constants,
the running behavior being determined by the corresponding $\beta$--functions. 
In the same way as the contact terms cannot consistently absorb higher order
divergences, their $\beta$--functions cannot compensate for scale dependence
which is suppressed by powers of $\mu$. 
In order to avoid this unphysical scale dependence in physical results, 
the authors of \cite{Becher} have argued that the nucleon mass $m_N$ serves as
a ``natural'' scale in a relativistic baryon ChPT loop calculation and that
therefore one should set $\lambda = m_N$ everywhere when using the infrared
regularization scheme. This was already suggested in \cite{BKKM} for the
framework of a relativistic theory with ordinary dimensional regularization. 
We will follow this idea, hence in our formulae, what would actually
be $\log (M_\pi/\lambda)$ will always be spelled out as $\log \mu$.

\medskip \noindent 
In the following calculation, we frequently compare our results to the equivalent
heavy baryon ones. We wish to emphasize that one can always regain the heavy baryon 
result from the infrared regularized relativistic one by performing a strict chiral
expansion of all involved loop functions. As a simple example, we again show the expansion
of the nucleon mass up to third order corresponding to eq.~(\ref{massrel}), but this
time using infrared regularization. The result is
\beq
m_N-m = -4 c_1 M^2 +\frac{3g_A^2}{2F^2} mM^2 I(m^2) ~,
\eeq
where the loop integral $I(m^2)$ is given by
\beq
I(m^2) = -\mu^2 \biggl(L+\frac{1}{16\pi^2}\log\mu\biggr)  
  +\frac{\mu}{16\pi^2}\biggl\{ \frac{\mu}{2}-\sqrt{4-\mu^2}  
  \arccos\biggl(-\frac{\mu}{2}\biggr) \biggr\} ~.
\eeq
(Definitions and results for this and all other loop functions needed in this work
can be found in appendix~\ref{app:loopfkt}.) Expanding to leading order, one finds
\beq
I(m^2) = -\frac{\mu}{16\pi} + {\cal O}(\mu^2) ~,
\eeq
which leads to the well--known result
\beq
m_N-m = -4 c_1 M^2 -\frac{3g_A^2M^3}{32\pi F^2} +{\cal O}(\mu^4)~,
\eeq
yielding a contribution non--analytic in the quark masses. As explained before, 
the loop function $I(m^2)$ contains a non--leading divergence which cannot be
absorbed to this order, but will be by an appropriate contact term at fourth order.
The calculation of the nucleon mass to fourth order and the reduction to the
heavy baryon limit is done in detail in \cite{Becher}. 

\medskip \noindent  
An essential ingredient to the treatment of loop integrals as described in the  
previous paragraphs is the fact that higher order effects are included as compared to the ``strict''  
chiral expansion in the heavy baryon formalism. This was justified in \cite{Becher} by improved
convergence properties in the low energy region, it introduces however a certain amount of  
arbitrariness as to which of these higher order terms to keep and which to dismiss.   
It is therefore mandatory to exactly describe our treatment of these terms. The  
philosophy in \cite{Becher} was, above all, to preserve the correct relativistic analyticity  
properties. This was achieved by keeping the full \emph{denominators} of loop integrals   
(and evaluating them by the infrared regularization prescription), while expanding the   
\emph{numerators} to the desired chiral order only. In addition, e.g.\ crossing symmetry   
is to be conserved. We explore a different approach here: as we anticipate the neutron   
electric form factor to be highly sensitive to recoil effects, we keep \emph{all} terms  
which occur according to the infrared regularization prescription and do not even expand   
the numerators of loop integrals. However, three effects involving low energy constants  
have to be discussed separately, all in the spirit of avoiding ``artificial'' higher  
order counterterms:  
\begin{itemize}  
\item Loop diagrams with second order insertions proportional to $c_1$ can easily be  
summarized by a shift of the ``bare'' nucleon mass to its renormalized value at second order,  
$m \rightarrow m_N = m-4c_1 M_\pi^2+{\cal O}(q^3)$. In principle, this does \emph{not} occur in   
those loop diagrams which have insertions from other second order low energy constants.  
As we do not want to artificially introduce this additional LEC by discriminating   
between $m$ and $m_N$ in our formulae, we allow for this renormalization everywhere.
\item As detailed in the previous section, 
the second order LECs $c_6,\,c_7$ receive a renormalization
$\sim M_\pi^2$ at fourth order. In principle, to this order only the unrenormalized 
$c_i$ appear in loop diagrams. We disregard this difference and use the same 
values both on tree level and for the loops.
\item From the definitions of the various electromagnetic form factors, it is obvious  
that these are not all calculated to the same accuracy in terms of chiral orders. A   
calculation employing chiral Lagrangians up to and including ${\cal O}(q^4)$ will  
be able to give the Dirac form factor $F_1$ to ${\cal O}(q^3)$ and the Pauli form factor  
$F_2$ to ${\cal O}(q^2)$. When combining these to the Sachs form factors $G_E$ and $G_M$,  
the chiral orders are ``mixed'', see eq.~(\ref{sachsdef}).   
When we, in the following, talk about ``third''/``fourth'' order calculations of 
the various form factors, we always refer to the projections of the third/fourth
order \emph{amplitude} onto these observables.
We truncate exactly one kind of term in this process: the ${\cal O}(q^3)$   
counterterms which enter the electric (charge) radii are really ``$G_E$ counterterms''  
in the sense that they appear in $F_1$ and $F_2$ with opposite signs in order to cancel   
exactly in $G_M$. In addition, there are, at ${\cal O}(q^4)$, $F_2$ counterterms fixing  
the magnetic radii.   
Therefore, $G_E$ receives  $q^4$ counterterms inherited  
from $({q^2}/{4m^2})F_2$. As a genuine polynomial contribution of this kind only   
appears in an ${\cal O}(q^5)$ calculation, we drop these terms in $G_E$.  
\end{itemize}  
 
 
\subsection{Chiral expansion of the nucleon form factors}  
\label{sec:chiexpff}  

The chiral expansion of a form factor $F$ (being a genuine symbol for any of the   
four electromagnetic nucleon form factors) consists of two contributions,   
tree and loop graphs. The tree graphs comprise that of lowest order with a   
fixed coupling (the nucleon charge) as well as counterterms from the second,  
third, and fourth order Lagrangians. As one--loop graphs, we have both those  
with just lowest order couplings and those with exactly one insertion from   
${\cal L}_{\pi N}^{(2)}$.  The pertinent tree and loop graphs are
depicted in fig.~\ref{fig:diag} (we have not shown the diagrams leading 
to wave function renormalization). The first four of these comprise the
tree graphs with insertions up to dimension four, and  diagrams (5) to (9) ((10)
to (12)) are the third (fourth) order loop graphs. Consequently, any
form factor can be written as
\beq 
F(t) = F^{\rm tree} (t) + F^{\rm loop} (t)~.
\eeq
In appendix~\ref{app:contri}, we have listed the contributions to the
Dirac and Pauli form factors from the various diagrams, including the
nucleon Z--factor. This allows to reconstruct the pertinent
expressions for $F_{1,2}^{p,n}$ in a straightforward manner. We
refrain from giving the complete expressions here. The corresponding Sachs
form factors $G_{E,M}^{p,n}$ are obtained by use of
eq.~(\ref{sachsdef}) under the restrictions discussed in the
preceding section.
We only show the explicit formulae for the magnetic moments, the electric
and the magnetic radii (in terms of renormalized, i.e.\ physical, masses and 
coupling constants). The magnetic moments read 
\beqa 
\kappa^v &=& \tilde{c}_6^r  \no\\  
&-&  \frac{\mu^2 g_A^2 m_N^2}{(4\pi F_\pi)^2} \biggl\{ 
 \frac{3}{2} \mu^2 +\biggl( 2-\frac{\mu^2}{4} \biggr)\, c_6  \no\\ 
&& \qquad
+ \,\frac{\arccos(-\frac{\mu}{2})}{\mu \sqrt{4-\mu^2}} 
\biggl[ 2 \Bigl( 8- 13\mu^2 + 3\mu^4 \Bigr) 
- \mu^2 \biggl( 8 - \frac{5}{2} \mu^2 \biggr)\, c_6 \biggr] \no\\  
&& \qquad
+ \,  \biggl[ 2 \Bigl(7-3\mu^2 \Bigr) + \biggl(4-\frac{5}{2}\mu^2\biggr)\, c_6 \biggr] \log \mu  
\biggr\} \no\\ 
&+&\frac{\mu^2 m_N^2}{(4\pi F_\pi)^2} \biggl\{ 
\frac{3}{4} \mu^2 m_N\,c_2 
- \Bigl( 3\mu^2 m_N\,c_2 -8m_N\,c_4 +2c_6 \Bigr) \log \mu \biggr\}~, \label{kappav} \\ 
\kappa^s &=& \tilde{c}_6^r + 2\tilde{c}_7^r \no\\ 
&-&  \frac{3 \mu^2 g_A^2 m_N^2}{(4\pi F_\pi)^2} \biggl\{ 
\frac{\mu^2}{4} \Bigl(  2+  c_6 +2 c_7 \Bigr) \no\\ 
&& \qquad
- \,\frac{\mu \arccos(-\frac{\mu}{2})}{\sqrt{4-\mu^2}} 
\biggl[ 2 \Bigl( 3 - \mu^2 \Bigr) 
+        \biggl( 4 - \frac{3}{2}\mu^2 \biggr) \Bigl( c_6 +2c_7 \Bigr) \biggr] \no\\ 
&& \qquad
+ \, \biggl[ 2 \Bigl( 1-\mu^2 \Bigr) 
   + \Bigl(2-\frac{3}{2}\mu^2\Bigr) \Bigl( c_6 +2c_7 \Bigr) \biggr] \log \mu  
\biggr\} \no\\ 
&+&\frac{3\mu^4 m_N^3\, c_2}{(4\pi F_\pi)^2} \biggl( \frac{1}{4}- \log \mu \biggr)~, 
\label{kappas}
\eeqa 
with $\mu = M_\pi /m_N \simeq 1/7$. Note that the magnetic moments are finite in the chiral limit.
We remind the reader of the definitions of $\tilde{c}_6$ and $\tilde{c}_7$ in
eq.~(\ref{ctilde}). We have distinguished between the renormalized and the
unrenormalized LECs (at tree level and in loop contributions, respectively)
only for reasons of formal exactitude, and will disregard this
difference from now on. 
Note however that, in contrast to $c_6,\,c_7$, these renormalized LECs 
inherit infinite parts from $e_{105},\,e_{106}$ which are needed in order to remove
the leading order divergences in the magnetic moments 
(as explained in section~\ref{sec:IR}). 
Their poles in $(d-4)$ are determined by the corresponding $\beta$--functions,
\beqa
\beta_{e_{105}} &=& -\frac{3g_A^2}{16m_N}\Bigl(1+c_6+2c_7\Bigr) ~,\\
\beta_{e_{106}} &=& -\frac{1}{8m_N} \Bigl[7g_A^2-4m_Nc_4+(1+2g_A^2)c_6 \Bigr] ~.
\eeqa 
$\tilde{c}_6^r,\,\tilde{c}_7^r$ in eqs.~(\ref{kappav}),~(\ref{kappas}) 
denote the finite LECs, i.e.\ with the poles subtracted.
Next, the charge radii can be brought into the form
\beqa
(r_E^v)^2 &=& -12 \,d_6^r  +\frac{3\,\kappa^v}{2m_N^2}\no\\
&-& \frac{g_A^2}{(4\pi F_\pi)^2} \biggl\{
  7 -8 \mu^2 -\frac{3}{4}\mu^4 \Bigl(1-c_6\Bigr) \no\\
&&  \qquad
- \, \frac{\mu\arccos(-\frac{\mu}{2})}{\sqrt{4-\mu^2}} 
  \Bigl[ 70 - 74 \mu^2 + 15 \mu^4 +3\mu^2 \Bigl(3-\mu^2 \Bigr) c_6 \Bigr] \no\\
&& \qquad
+ \,\Bigl[ 10-44\mu^2+15\mu^4+3\mu^2\Bigl(1-\mu^2\Bigr) c_6 \Bigr] \log\mu
  \biggr\} \no\\
&-& \frac{1}{(4\pi F_\pi)^2} \Bigl( 1 + 2\log\mu \Bigr) ~, \label{rEv2} \\
(r_E^s)^2 &=& -24 \,d_7 + \frac{3\,\kappa^s}{2m_N^2} \no\\
&+& \frac{3\mu^2g_A^2}{(4\pi F_\pi)^2} \biggl\{
\frac{2}{4-\mu^2} +\frac{\mu^2}{4} \Bigl[1+3\Bigl(c_6+2c_7\Bigr) \Bigr]  \no\\
&& \qquad
- \,\frac{\mu\arccos(-\frac{\mu}{2})}{(4-\mu^2)^{3/2}} 
  \Bigl[ 54 - 34 \mu^2 + 5 \mu^4  
  + 3 \Bigl(12-7\mu^2+\mu^4\Bigr) \Bigl(c_6+2c_7\Bigr) \Bigr] \no\\
&& \qquad
+ \, \Bigl[ 4-5\mu^2+3\Bigl(1-\mu^2\Bigr)\Bigl(c_6+2c_7\Bigr)\Bigr] \log\mu
  \biggr\} ~.
\label{rEs2}
\eeqa
The isovector charge radius diverges logarithmically in the chiral limit, 
while the isoscalar one is finite. $d_6^r$ denotes the renormalized LEC $d_6$, i.e.\ 
without  the infinite part that removes the leading order divergence
in the isovector electric form factor,
the corresponding $\beta$--function being given by \cite{GE,FMS}
\beq
\beta_{d_6}=-\frac{1+5g_A^2}{6} ~.
\eeq 
The LEC $d_7$ entering the isoscalar radius is, in contrast, finite.
Finally, the magnetic radii are given by
\beqa
\mu^v (r_M^v)^2 &=& 24 \,m_N \, e_{74}^r \no\\
&+&\frac{g_A^2}{(4\pi F_\pi)^2} \biggl\{
\frac{1}{4(4-\mu^2)} \Bigl[ 112-36\mu^2-8\mu^4+3\mu^6 
  -\mu^2 \Bigl( 8-8\mu^2+\mu^4\Bigr) \,c_6  \Bigr]\no\\
&& \qquad+ \,\frac{\arccos(-\frac{\mu}{2})}{\mu (4-\mu^2)^{3/2}} 
  \Bigl[ 32-152\mu^2+126\mu^4-34\mu^6+3\mu^8
 -\mu^4\Bigl( 6-6\mu^2+\mu^4 \Bigr) \,c_6 \Bigr]\no\\
&& \qquad + \,\Bigl[ 14 - 16\mu^2 + 3\mu^4 \Bigl( 3 - c_6 \Bigr) \Bigr] \log \mu
\biggr\} \no\\
&-&\frac{1}{(4\pi F_\pi)^2}\Bigl( 1 + 2\log \mu \Bigr) \Bigl(1 +4m_N\,c_4 \Bigr)
~, \label{rMv2} \\
\mu^s (r_M^s)^2 &=& 48 \,m_N \, e_{54}  \no\\
&-&\frac{3\mu^2g_A^2}{(4\pi F_\pi)^2} \biggl\{
\frac{8-8\mu^2+\mu^4}{4(4-\mu^2)}
+\frac{\mu\arccos(-\frac{\mu}{2})}{(4-\mu^2)^{3/2}}\Bigl( 6-6\mu^2+\mu^4 \Bigr)
+\mu^2 \log\mu \biggr\} \no\\
&& \qquad \qquad \times \Bigl(1+c_6+2c_7 \Bigr)
~, \label{rMs2}
\eeqa
where, again, $e_{74}^r$ has the infinite part, which is given by
\beq
\beta_{e_{74}} = -\frac{1}{3m_N}\Bigl(3g_A^2-m_Nc_4  \Bigr) ~,
\eeq
subtracted, while $e_{54}$ in the isoscalar radius 
(in analogy to $d_7$ for the electric isoscalar radius) is finite.
The isovector magnetic radius  explodes like $1/M_\pi$ as the pion mass goes to zero,
while, again, the isoscalar radius is finite in the chiral limit.
These chiral limit results are
to be expected because the Sachs form factors inherit the well known chiral singularities
of the isovector Dirac and Pauli form factors,
more precisely, the corresponding radii diverge in the chiral limit as 
\beqa\label{chilimrad} 
(r_1^v)^2 &=& - {10 g_A^2 + 2\over (4\pi F_\pi)^2}\,\log\frac{M_\pi}{m_N} + 
\ldots ~, \no \\ 
(r_2^v)^2 &=& { g_A^2 m_N \over 8\pi F_\pi^2 \kappa_v M_\pi} + \ldots ~, 
\eeqa 
where the ellipsis denotes subleading terms as the pion mass vanishes. This dominant  
pion loop effect in the nucleon form factors has already been established a long 
time ago~\cite{BZ}. It can also be expected on physical grounds: the pion
cloud, which is Yukawa suppressed for finite pion mass, becomes long--ranged in
the chiral limit leading to a divergent contribution. Such phenomena are also
observed in two--photon observables like e.g.\ the electromagnetic
polarizabilities of the nucleon.
Also, as a consequence of analyticity, in the heavy baryon approach (i.e.\ 
in a strict chiral expansion) the isoscalar form factors are, to one--loop 
order, given simply by a polynomial subsuming tree and counterterm contributions.
There \emph{are}, however, isoscalar loop effects as higher order corrections,
as is obvious from the above. These stem from the diagrams with the photon coupling
to the nucleons (see graphs (5) and (10) in fig.~\ref{fig:diag}), which only
yield constant terms needed for renormalization in HBChPT, but also contain higher order
$t$--dependent terms. As the spectral function corresponding to these diagrams
only starts to contribute for $t \geq 4m_N^2$, these $t$--dependent terms (in the real
part) are highly suppressed.

\medskip \noindent
In section~\ref{sec:IR}, we already mentioned the marked difference in the threshold
behavior of the isovector spectral functions in the relativistic and the heavy
baryon approach. We only quote the results for the threshold behavior here
and relegate the full results to appendix~\ref{app:spectral}.
Note that these have already been given
in the literature, the relativistic results to third order in \cite{GSS}, the
additional fourth order contribution as well as the heavy baryon expansions
in \cite{spectral}. We reemphasize that the imaginary parts in the infrared 
regularization approach do not differ from those in a fully relativistic calculation.
For Im$\,F_1^v(t)$, the relativistic threshold expansion is given by
\beq
{\rm Im}\,F_1^v(t) = 
\frac{1}{192\pi F_\pi^2 \,M_\pi^3} \biggl\{
  g_A^2 \Bigl(4m_N^2-M_\pi^2\Bigr) + M_\pi^2 \biggr\} \, (t-4M_\pi^2)^{3/2}
+ {\cal O}\Bigl((t-4M_\pi^2)^{5/2}\Bigr)~,
\eeq
and for Im$\,F_2^v(t)$, one finds
\beq
{\rm Im}\,F_2^v(t) = 
\frac{m_N\,c_4}{48\pi F_\pi^2\,M_\pi} \, (t-4M_\pi^2)^{3/2}
+{\cal O}\Bigl((t-4M_\pi^2)^{5/2}\Bigr)
~.
\eeq
In the heavy baryon scheme, however, the heavy mass expansion yields the expressions
\beqa
{\rm Im} \, F_1^v(t) &=&
\frac{g_A^2}{96\pi F_\pi^2} \biggl\{ \sqrt{1-\frac{4M_\pi^2}{t}} 
  \Bigl( 5t - 8M_\pi^2 \Bigr) 
  - \frac{3\pi}{4m_N}\,\frac{3t^2-12M_\pi^2t+8M_\pi^4}{t^{1/2}} \biggr\} \no\\
&& +\,\frac{1}{96\pi F_\pi^2} \, \frac{(t-4M_\pi^2)^{3/2}}{t^{1/2}} \,
+{\cal O}(q^4) ~, \\
{\rm Im} \, F_2^v(t) &=&
\frac{g_A^2\,m_N}{32F_\pi^2} \biggl\{ \frac{t-4M_\pi^2}{t^{1/2}}
- \frac{4}{\pi m_N} \,\sqrt{1-\frac{4M_\pi^2}{t}} \,\Bigl( t-2M_\pi^2 \Bigr) \biggr\}\no\\
&&+\,\frac{m_N\,c_4}{24\pi F_\pi^2} \, \frac{(t-4M_\pi^2)^{3/2}}{t^{1/2}} \, 
+{\cal O}(q^3) ~,
\eeqa
which evidently violate the required threshold behavior 
${\rm Im} \, F_i^v(t) \sim (t-4M_\pi^2)^{3/2}$ as pointed out in
\cite{spectral}. As remarked before, this is due to the fact that
the two--pion threshold and the singularity on the second Riemann
sheet, inherited from the $\pi\pi \to \bar{N}N$ partial waves,
coalesce in the heavy baryon limit. In contrast, the relativistic
approach of course yields the correct analytical structures (for a
more pedestrian discussion of this point, see e.g.\ \cite{Ulfhugs}). 
  
\subsection{Results and discussion}  
\label{sec:res}  
  
First, we must fix parameters.  We use $F_\pi = 92.4\,$MeV, $M_\pi = 139.57\,$MeV, 
$g_A = 1.26$, and $m_N = (m_p + m_n)/2 = 938.92\,$MeV. The LECs $c_2$ and $c_4$ 
are taken from the analysis of pion--nucleon scattering \cite{FMS,Paul}, 
$c_2 = 3.2\,$GeV$^{-1}$ 
and $c_4 = 3.4\,$GeV$^{-1}$. The LECs $c_6$ and $c_7$ can be pinned down from 
the well known magnetic moments of proton and neutron. The remaining 
two ``electric'' ($d_6, \, d_7)$ and two ``magnetic'' LECs ($e_{54},\,e_{74})$ are  
determined from the electric and 
magnetic radii as given by the dispersion theoretical analysis of \cite{mmd}, these 
values are $r_E^p = 0.847\,$fm,  $(r_E^n)^2 = -0.113\,$fm$^2$,  $r_M^p = 0.836\,$fm 
and  $r_M^n = 0.889\,$fm. We note that the long standing discrepancy of the proton 
charge radius determination from electron--proton scattering \cite{RR} 
and Lamb shift measurements \cite{MvR} 
has been resolved, converging to a value of $r_E^p = (0.88\pm 0.01)\,$fm. We do not use 
this value here because so far no dispersion theoretical analysis exists using 
this novel information.  
 
\medskip \noindent 
As already stated, the Dirac and Pauli form factors are the natural quantities 
in any relativistic approach like the one used here. However, for easier comparison to
existing heavy baryon calculations and to the low energy data, which are  
often presented in terms of Sachs form factors, we will discuss these in detail. 

\begin{figure}[htb] 
\centerline{ 
\epsfysize=4.2in 
\epsffile{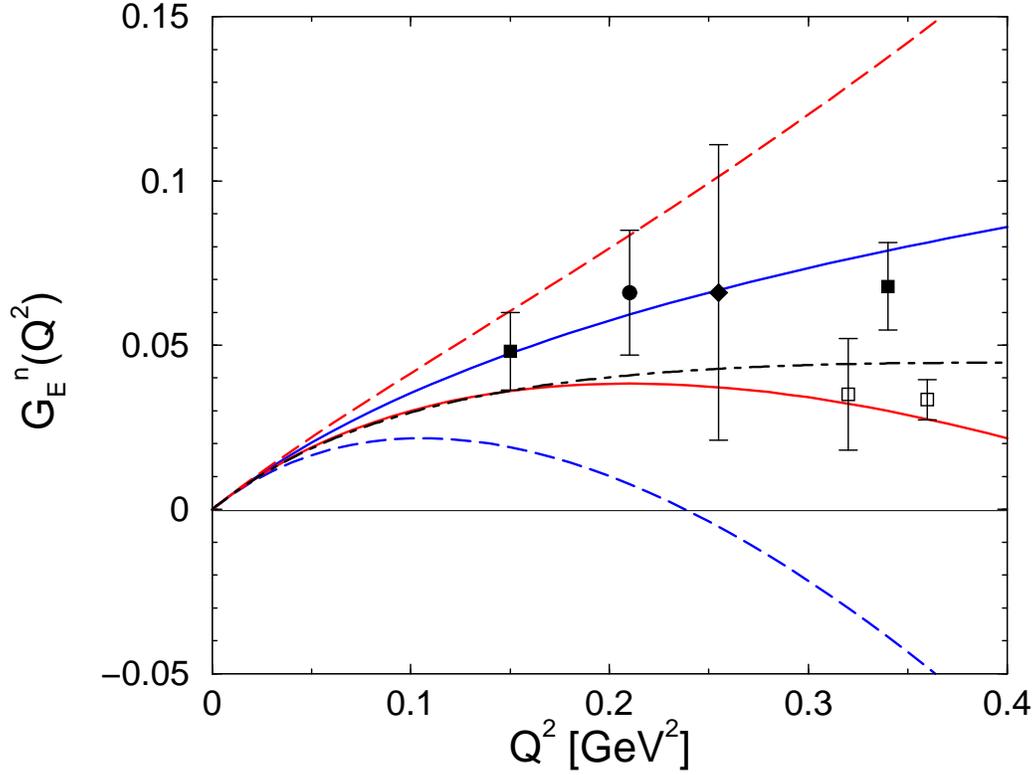} 
} 
\vskip 2.5cm 
\caption{The neutron electric form factor in relativistic baryon chiral 
perturbation theory (solid lines) to third (blue curve) and fourth (red 
curve) order. For comparison, the results of the heavy baryon approach are 
also shown (blue/red dashed line: third/fourth order). All LECs are 
determined by a fit to the neutron charge radius measured in  
neutron--atom scattering. Also given is the result of  the dispersion 
theoretical analysis (black dot--dashed curve). The data are from 
\protect\cite{newnff}. 
\label{fig:GEn} 
} 
\end{figure}

\medskip \noindent 
First we discuss  the neutron charge form factor, as shown in fig.~\ref{fig:GEn}. 
The third and fourth order results compare well 
to the dispersion theoretical analysis and the trend of the novel data up to  
$Q^2 = 0.4\,$GeV$^2$. 
The latter have been obtained using tensor polarized deuterium, 
polarized $^3$He and polarization transfer on deuterium~\cite{newnff}. 
Although it seems that the \emph{third} order curve meets the
novel data even better, we observe that the fourth order one is basically
identical to the dispersion theoretical fit up to $Q^2 = 0.2\,$GeV$^2$.
Also given in that figure are the heavy baryon results to third \cite{BFHM}  
and fourth \cite{HG} order. 
These curves can easily be obtained from our analysis by performing the large mass 
expansion as explained before.
Clearly, neither of these two curves is in acceptable agreement with the data, 
and, what is more, there is no sign of convergence so far.
The resummation of the $1/m$ terms in the relativistic approach 
vastly improves the convergence of the chiral representation.
This sensitivity of the neutron charge form factor to such recoil corrections was 
already anticipated in~\cite{KHM}. 
We remark however that the only fourth order contributions to the electric form factors
in the heavy baryon approach are $1/m$--corrections to third order diagrams, 
in other words, the full fourth order heavy baryon result for these
can already be obtained from expanding the \emph{third} order relativistic one. 
Therefore, the difference that does show up between the 
relativistic third and fourth order curves is really, in terms of strict chiral
power counting, of higher order and hence (comparably) small.

\begin{figure}[htb] 
\centerline{ 
\epsfysize=4.2in 
\epsffile{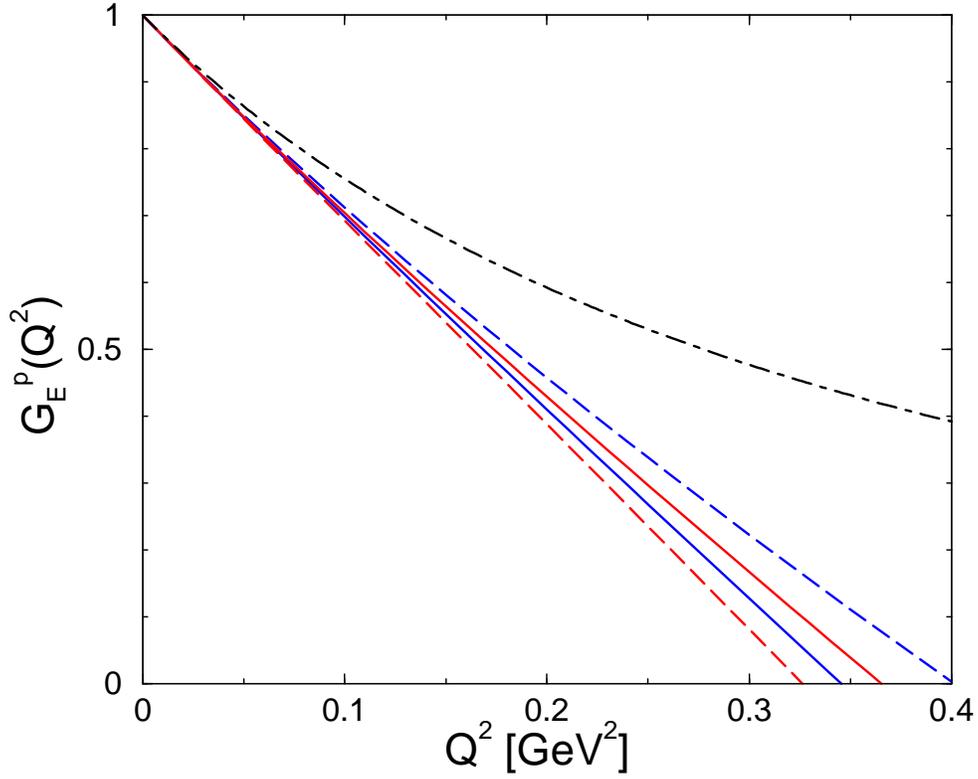} 
} 
\vskip 2.5cm 
\caption{The proton electric form factor in relativistic baryon chiral 
perturbation theory (solid lines) to third (blue curve) and fourth (red 
curve) order. For comparison, the results of the heavy baryon approach are 
also shown (blue/red dashed line: third/fourth order). All LECs are 
determined by a fit to the proton charge radius as given by the dispersion 
theoretical result (black dot--dashed curve). 
\label{fig:GEp} 
} 
\end{figure}

\medskip \noindent 
Consider the proton electric form factor next, shown in fig.~\ref{fig:GEp}. 
As in the neutron case, the third and fourth order curves are very close
(where, however, the remarks made about the reasons for the smallness 
of this difference apply equally here), 
but here essentially show the linear behavior from the term 
proportional to the charge radius. Compared to the dispersion theoretical 
result (which describes the data very well in this range of momentum transfer), 
there is clearly not enough curvature.
Polynomial terms of order $t^2$ (and higher) are not included up to fourth order,
and the curvature coming solely from loop functions in the one--loop approximation 
is obviously not sufficient.
Fig.~\ref{fig:GEp} also displays the results of the third~\cite{BKKM,BFHM} and   
fourth~\cite{HG} order heavy baryon calculation.  
While the general trend of the heavy baryon results is similar to what is obtained in 
the relativistic approach (far too small curvature to meet the data), 
we note that, while there is at least a slight improvement from the third to the fourth
order relativistic calculation, the description visibly \emph{worsens} for the
fourth order heavy baryon result, as compared to the third order one.

\begin{figure}[htb] 
\centerline{ 
\epsfysize=4.2in 
\epsffile{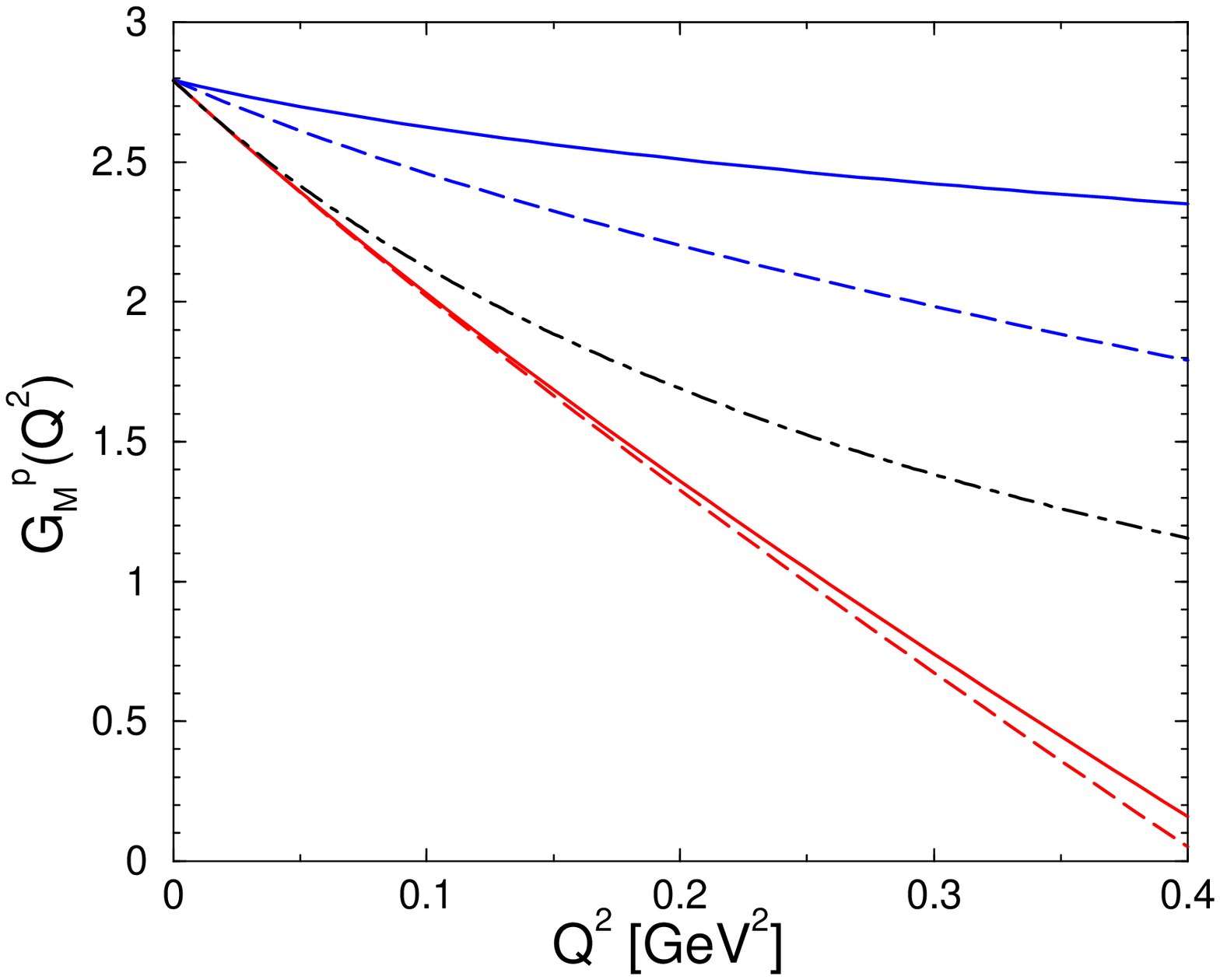} 
} 
\vskip 2.5cm 
\caption{The proton magnetic form factor in relativistic baryon chiral 
perturbation theory (solid lines) to third (blue curve) and fourth (red 
curve) order. For comparison, the results of the heavy baryon approach are 
also shown (blue/red dashed line: third/fourth order). The third (fourth) order LECs are 
determined by a fit to the proton magnetic moment (radius). Also given is the dispersion 
theoretical result (black dot--dashed curve). 
\label{fig:GMp} 
} 
\end{figure}

\begin{figure}[htb] 
\centerline{ 
\epsfysize=4.2in 
\epsffile{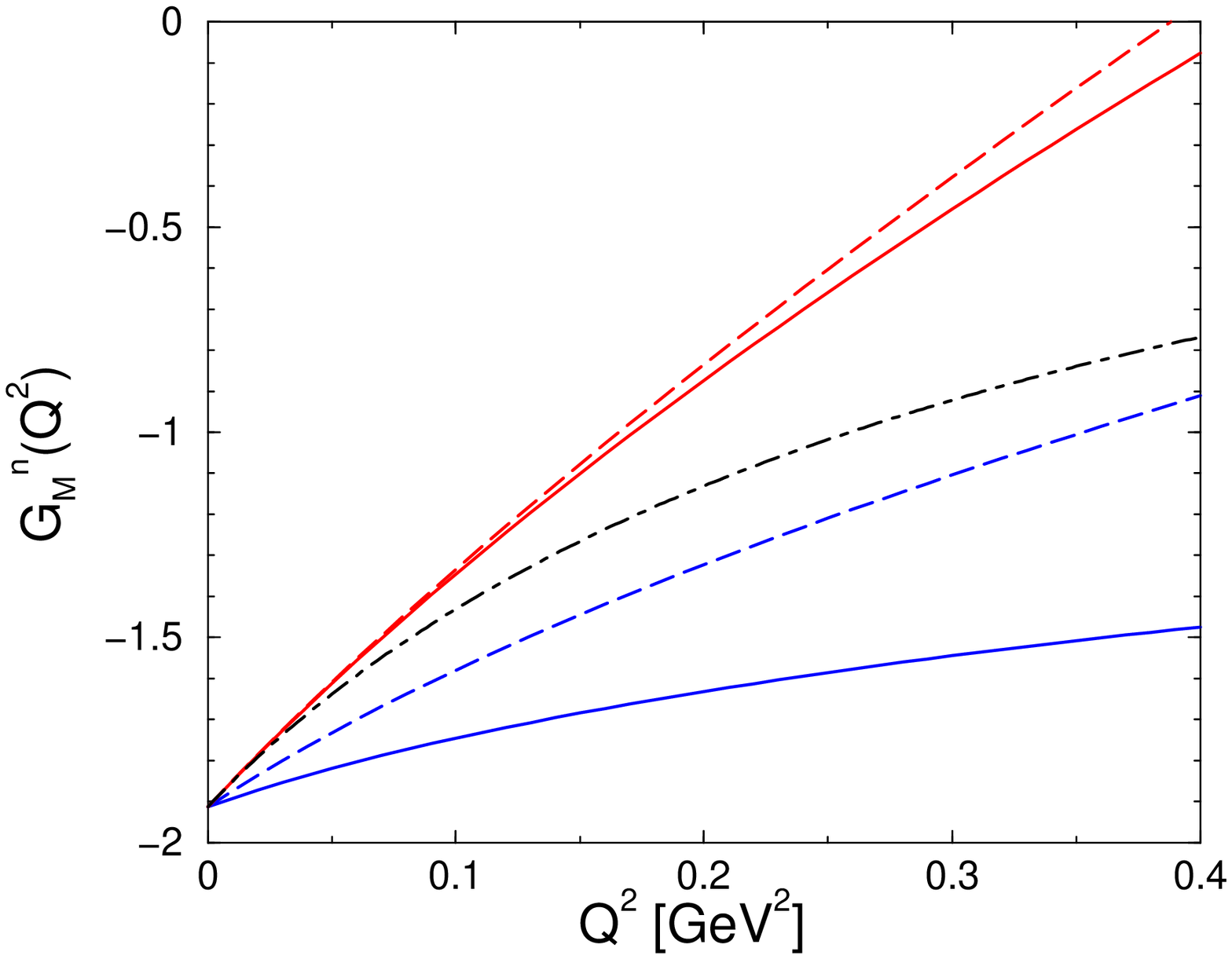} 
} 
\vskip 2.5cm 
\caption{The neutron magnetic form factor in relativistic baryon chiral 
perturbation theory (solid lines) to third (blue curve) and fourth (red 
curve) order. For comparison, the results of the heavy baryon approach are 
also shown (blue/red dashed line: third/fourth order). The third (fourth) order LECs are 
determined by a fit to the neutron magnetic moment (radius). Also given is the dispersion 
theoretical result (black dot--dashed curve). 
\label{fig:GMn} 
} 
\end{figure}

\medskip \noindent
We now turn to the magnetic form factors of proton and neutron as shown in 
figs.~\ref{fig:GMp},~\ref{fig:GMn}. The qualitative behavior of the different curves
is very similar for both of them:
To third order, the momentum dependence is given parameter--free 
because the LECs at this order only affect the normalization. 
We see that the $1/m$ corrections present in our approach visibly worsen the 
prediction obtained previously in the heavy baryon limit. That result is based 
on the leading chiral singularities given in eq.~(\ref{chilimrad}) because to this order, 
the corresponding isoscalar form factor is simply constant. 
We conclude that the recoil corrections are rather large
and that the leading chiral limit behavior is not a good approximation 
for the Goldstone boson contribution to the magnetic radii in the case of finite pion masses. 
At fourth order, due to 
the presence of the counterterms from ${\cal L}_{\pi N}^{(4)}$, the radius can be 
fixed at its empirical value. Consequently, the full relativistic result and its heavy 
baryon limit are not very different any more. 
The absence of curvature terms of order $t^2$ (and higher) is clearly visible 
in  figs.~\ref{fig:GMp},~\ref{fig:GMn}. 

\medskip \noindent
The physics underlying this deficiency will be addressed in section~\ref{sec:vectors}. 
Only a uniformly reliable description for all four form factors is acceptable, and hence
it is absolutely necessary to understand what is missing in the
three dipole--shaped form factors: with large corrections for these, one might expect
large corrections also for the neutron charge form factor, such that its very good
description presented here might turn out to be only accidental. 
In the following, we shall show that this is indeed \emph{not} the case.
 
\section{Inclusion of vector mesons}  
\label{sec:vectors}  
\def\theequation{\arabic{section}.\arabic{equation}}  
\setcounter{equation}{0}  

It is well established that the vector mesons contribute significantly 
to the nucleon form factors. For example, extended unitarity allows 
one to reconstruct the isovector spectral functions~\cite{FraFu} below $t \simeq 
1\,$GeV$^2$ from the pion form factor and the analytically continued $\pi \pi \to 
\bar N N$ isovector partial wave amplitudes. Besides the important 
contribution of the two--pion continuum, the $\rho$ meson clearly shows 
up. Similarly, in the isoscalar channel, the $\omega$ and the $\phi$ 
dominate the spectral function at low positive $t$. All these effects 
are clearly visible in  dispersion theoretical analyses of the 
nucleon form factors, see~\cite{hoehler,mmd,hmd}. In the chiral 
expansion performed in the preceding sections, such effects are included in the 
low energy constants. This follows directly from the low momentum 
expansion of any vector meson propagator (for simplicity, we do not 
show the explicit Lorentz structures), 
\beq\label{Vprop} 
\dfrac{1}{1- {t}/{M_V^2}} = 1 + \dfrac{t}{M_V^2} +\dfrac{t^2}{M_V^4} + 
{\cal O}(t^3)~. 
\eeq  
In a fourth order chiral analysis as presented here, one is 
sensitive up to the terms linear in $t$, which contribute to the various 
electromagnetic radii. Stated differently, any vector meson 
contribution is hidden in the fit values of the various LECs. As we 
discussed before, the curvature effects on the  electric and 
magnetic form factor of the proton as well as the 
magnetic neutron form factor are much too small. This can be cured in 
two ways. One could either attempt a higher order analysis or include 
vector mesons dynamically. While the first approach would be more 
systematic, we choose here the second, for  various reasons. With 
explicit vector mesons (VM), we not only account for all terms in the 
expansion eq.~(\ref{Vprop}) but also do not introduce any new unknown 
parameter. This can be understood as follows: adding the vector mesons 
in a chirally symmetric manner and retaining the corresponding 
dimension two, three and four counterterms, any LEC $\alpha_i$ takes the form 
\beq 
\alpha_i \,\to\, \hat{\alpha}_i \,+\, {\rm vector~meson}\!-\!{\rm contribution}~, 
\eeq 
where the remainder $\hat{\alpha}_i$ parameterizes the physics not 
related to the explicitly included vector meson contribution.
For such a decomposition to make sense, the scale dependent LECs should be 
taken at $\lambda = M_V$. Since in the infrared regularization procedure,
the nucleon mass is taken to be the intrinsic scale for all loop integrals
as described in section~\ref{sec:IR}, here we neglect scale mismatch due 
to $M_V \neq m_N$ (which is probably justified because of $\log (m_N/M_\rho) \approx 0.2$).
The vector meson propagator generates a whole string of higher order terms, hopefully resumming 
the most important contributions not included at fourth order. The 
parameters appearing in the vector meson contributions will be taken 
from existing dispersive analyses of the nucleon electromagnetic 
form factors. We therefore only need to refit the low energy constants $\hat{\alpha}_i$. 
This also allows to study the concept of resonance saturation. 
It was already shown in~\cite{BKMpin} that the numerical values 
of the dimension two LECs can be understood from s--channel baryon 
and t--channel meson resonance excitations, in particular from the 
$\Delta (1232)$ and the $\rho$. In the case at hand, we can 
investigate this concept for the LECs $c_{6}$, $c_{7}$, $d_{6}$, $d_{7}$,  
$e_{54}$, and $e_{74}$. It was already noted in~\cite{BKMpin} that $c_6$ and 
$c_7$ are largely saturated by vector meson contributions.  
 

\subsection{Chiral Lagrangians including vector mesons}  
\label{sec:tensor} 
We employ the tensor field representation of spin-1 fields as advocated in \cite{GLAnn},   
i.e.\ the vector mesons are written in terms of antisymmetric tensor fields  
\beq  
W_{\mu\nu}=-W_{\nu\mu}  
\eeq  
with the three degrees of freedom $W_{ij}$ ($i,j=1,2,3$) frozen out. This 
representation is most natural for constructing chirally invariant couplings
of vector mesons to pions and photons because no particular dynamical 
character of the vector mesons is assumed (in contrast e.g.\ to the 
massive Yang--Mills or hidden symmetry approaches as reviewed in~\cite{UGM,koichi}.)   
In this section, we temporarily switch to SU(3) chiral Lagrangians, because the 
$\phi$ meson contributes to the isoscalar form factors.  
The free Lagrangian then takes the form  
\beq  
{\cal L}_V = -\frac{1}{2} \partial^\mu W^a_{\mu\nu} \partial_\rho W^{\rho\nu,a}  
+ \frac{M_V^2}{4}W^a_{\mu\nu} W^{\mu\nu,a}~,  
\label{VMLagr}
\eeq  
where  
\beq  
W_{\mu\nu} = \left(   
\begin{array}{ccc}  
\frac{\rho^0}{\sqrt{2}} + \frac{\omega}{\sqrt{2}} & \rho^+ & K^{*+} \\   
\rho^- & - \frac{\rho^0}{\sqrt{2}} + \frac{\omega}{\sqrt{2}} & K^{*0} \\   
K^{*-} & \overline{K}^{*0} & -\phi  
\end{array}  
\right)_{\mu\nu}~.  
\eeq  
Here, we have written the vector meson matrix in terms of physical particles, assuming  
ideal $\phi$--$\omega$ mixing.  From eq.~(\ref{VMLagr}), one easily derives the
vector meson propagator as given in \cite{Borasoy},
\beqa
G_{\mu\nu,\rho\sigma}(x,y) &=& \langle 0 | T\{W_{\mu\nu}(x),W_{\rho\sigma}(y)\} | 0\rangle \no\\
&=& \frac{i}{M_V^2} \int \frac{d^4k}{(2\pi)^4} \frac{e^{ik\cdot(x-y)}}{M_V^2-k^2-i\epsilon} \no\\
&& \times \Bigl[ g_{\mu\rho}g_{\nu\sigma}(M_V^2-k^2) +g_{\mu\rho}k_\nu k_\sigma
                 -g_{\mu\sigma}k_\nu k_\rho - (\mu \leftrightarrow \nu) \Bigr] ~.
\label{VMProp}
\eeqa
According to \cite{EGPR}, the lowest order interaction with Goldstone boson fields as  
well as external vector and axial--vector sources can be written as  
\beq  
{\cal L}_W =  
\frac{1}{2\sqrt{2}}\Bigl(   
F_V \langle W^{\mu\nu} F^+_{\mu\nu} \rangle   
+ i G_V \langle W^{\mu\nu} [ u_\mu, u_\nu ] \rangle \Bigr) ~, 
\label{VMpi}
\eeq 
where the couplings $G_V$ and $F_V$ can e.g.\ be determined from the decay widths 
$\rho\to \pi\pi$ and $\rho\to e^+e^-$.   
Following \cite{Borasoy}, the lowest order couplings of massive spin--1 fields to   
baryons can be written in terms of a chirally invariant Lagrangian as  
\beqa  
{\cal L}_{\phi B W} &=&   
R_{D/F} \, \langle \bar{B} \sigma^{\mu\nu} (W_{\mu\nu},B)_\pm \rangle \,  
+ \, R_S \, \langle \bar{B} \sigma^{\mu\nu} B \rangle \, \langle W_{\mu\nu} \rangle \no\\  
&& + S_{D/F} \, \langle \bar{B} \gamma^\mu ([D^\nu,W_{\mu\nu}],B)_\pm \rangle \,  
+\, S_S \,\langle \bar{B} \gamma^\mu B \rangle \, \langle [D^\nu,W_{\mu\nu}] \rangle \no\\  
&& + U_{D/F} \, \langle \bar{B} \sigma^{\lambda\nu}(W_{\mu\nu},[D_\lambda,[D^\mu,B]])_\pm \rangle \,  
+ \, U_S \, \langle \bar{B} \sigma^{\lambda\nu} [D_\lambda,[D^\mu,B]] \rangle \langle W_{\mu\nu} \rangle ~,  
\eeqa  
where, as usually in SU(3), the index $D$ refers to the anticommutator $(A,B)_+=\{A,B\}$,  
while the index $F$ accompanies the commutator $(A,B)_-=[A,B]$. 
In addition, we introduced singlet couplings with indices $S$. 
These coupling constants are related to the ones used in the Lagrangian  
of the standard vector representation of the $\rho$,   
\beq  
{\cal L}_{\rho N}=\frac{1}{2}g_{\rho NN} \bar{\Psi} \Bigl\{  
\gamma^\mu \boldsymbol{\rho}_\mu \! \cdot \! \boldsymbol{\tau}  
- \frac{\kappa_\rho}{2m_N}\sigma^{\mu\nu} \partial_\nu   
  \boldsymbol{\rho}_\mu \! \cdot \! \boldsymbol{\tau}  
\Bigr\} \Psi ~,  
\eeq  
by  
\beqa  
g_{\rho NN} &=& \frac{M_V}{\sqrt{2}} \Bigl( m_N(U_D+U_F) - 2(S_D+S_F) \Bigr) ~, \\  
g_{\rho NN} \, \kappa_\rho &=& - \frac{4\sqrt{2} m_N}{M_V} (R_D+R_F) ~.   
\eeqa  
In analogy, one finds for the $\omega$ and $\phi$ couplings
\beqa  
g_{\omega NN} &=& \frac{M_V}{\sqrt{2}} \Bigl( m_N(U_D+U_F+2U_S) - 2(S_D+S_F+2S_S) \Bigr) ~, \\  
g_{\omega NN} \, \kappa_\omega &=& - \frac{4\sqrt{2} m_N}{M_V} (R_D+R_F+2R_S) ~, \\  
g_{\phi NN} &=& - M_V  \Bigl( m_N(U_D-U_F+U_S) - 2(S_D-S_F+S_S) \Bigr) ~, \\  
g_{\phi NN} \, \kappa_\phi &=& \frac{8 m_N}{M_V} (R_D-R_F+R_S) ~.   
\eeqa  
In \cite{Borasoy}, one additional term (proportional to new couplings 
$T_{D/F}$) of higher order in derivatives was introduced,  
\beq  
{\cal L}'_{\phi B W} =  
T_{D/F} \, \langle \bar{B} \gamma^\mu ([D_\lambda,W_{\mu\nu}],  
[D^\lambda,[D^\nu,B]])_\pm \rangle \,  
+ \, T_S \, \langle \bar{B} \gamma^\mu [D^\lambda,[D^\nu,B]] \rangle \,   
\langle [D_\lambda,W_{\mu\nu}] \rangle ~,  
\label{Tterm}
\eeq  
which was subsequently shown to contribute to the electric form factors  
as a term ${\cal O}(q^4)$,  
therefore beyond the order to which we are working here; the authors of \cite{Borasoy}  
however left out a term  
\beq  
{\cal L}''_{\phi B W} =  
V_{D/F} \, \langle \bar{B} \sigma^{\nu\lambda}   
([D_\lambda,[D^\mu,W_{\mu\nu}]],B)_\pm \rangle \,  
+ \, V_S \, \langle \bar{B} \sigma^{\nu\lambda} B \rangle \,  
\langle [D_\lambda,[D^\mu,W_{\mu\nu}]] \rangle  
\eeq  
which enters the magnetic radius, i.e.\ contributes a term ${\cal O}(q^2)$ to the magnetic   
form factors and hence should be considered in a ${\cal O}(q^4)$ calculation.   
However, as nothing is known from elsewhere about such couplings, we shall   
not introduce these here.\footnote{The terms in eq.~(\ref{Tterm}) were allowed for 
in the analysis \cite{Oller} of $\pi N$--scattering and found to be very small. This lends
credit to the omission of higher order couplings here.}
Note that, via the $q^2$--expansion of the resonance pole, the contributions 
to the magnetic moments stemming from the $R_{D/F/S}$ terms \emph{do} of course induce   
additional $q^2$--dependence in the magnetic form factors.  
  
\begin{figure}[htb] 
\centerline{ 
\epsfysize=3.5in 
\epsffile{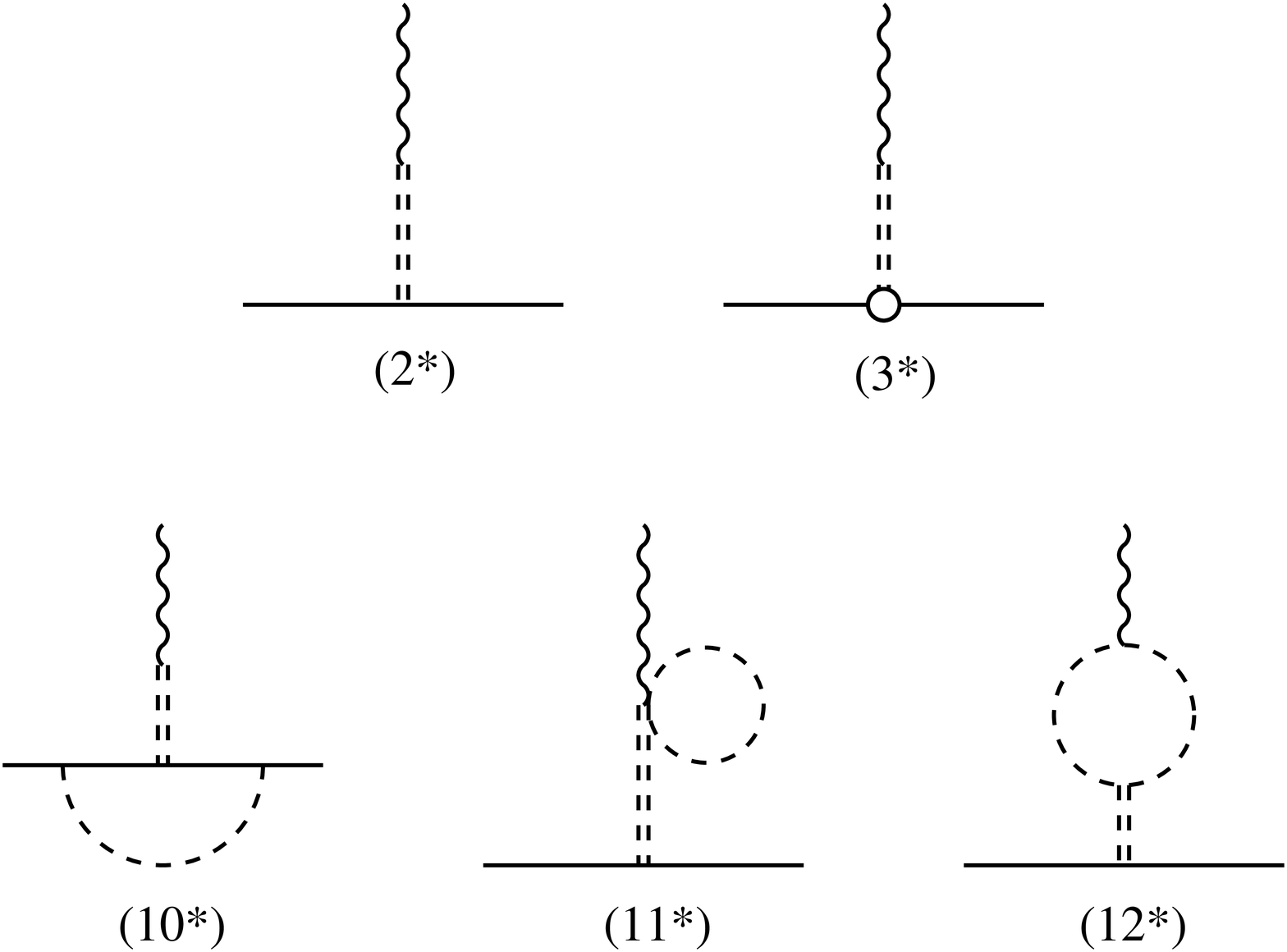} 
}
\vskip 3.5cm 
\caption{Feynman diagrams including explicit vector meson contributions
to the electromagnetic form factors up to fourth order. 
Solid, dashed, double--dashed, and wiggly lines refer to nucleons,
pions, vector mesons, and the vector source, respectively. 
The vertex denoted by an open dot refers to the vector coupling 
of the vector mesons to the nucleon
which is of subleading chiral order as compared to the tensor
coupling.
\label{fig:diagVM} 
} 
\end{figure}

\medskip \noindent  
There is no generalization of chiral perturbation theory which fully includes the
effects of vector mesons as intermediate states to arbitrary loop orders. As the masses
of the vector mesons do not vanish in the chiral limit, they introduce a new mass scale
which, when appearing inside loop integrals, potentially spoils chiral power counting 
in a similar manner as the nucleon mass in a ``naive'' relativistic baryon ChPT approach. 
In analogy to the heavy fermion theories, 
``heavy meson effective theory'' has been used to investigate vector meson properties
like masses and decay constants when coupling these to Goldstone bosons
in a chirally invariant fashion  (see \cite{BGT} or for some special processes \cite{JMW}).
This approach does not work with the heavy particle number not conserved 
as in processes with these resonances being virtual intermediate states. 
However, these difficulties do not arise as long as there is no loop integration 
over intermediate vector meson momenta.
Indeed, to the order we are working here, we can even set up a ``power counting scheme''
for diagrams including vector mesons which allows to calculate corrections to the
simple tree diagrams 
(where the photon couples to the nucleon via an intermediate vector meson).
We count
\begin{enumerate}
\item the couplings of vector mesons to the photon and of the $\rho$ to two pions as 
${\cal O}(q^2)$ (see eq.~(\ref{VMpi}) and the usual power counting for $F_{\mu\nu}^+$
and $u_\mu$);
\item the tensor coupling of vector mesons to nucleons ($R_{D/F/S}$ in our notation)
as ${\cal O}(q^0)$, the vector coupling ($S_{D/F/S},\,U_{D/F/S}$) as ${\cal O}(q^1)$;
\item the vector meson propagator as ${\cal O}(q^0)$ (see eq.~(\ref{VMProp})).
\end{enumerate}
Up to ${\cal O}(q^4)$, we therefore have to include the diagrams shown in 
fig.~\ref{fig:diagVM}. The numbering indicates the correspondence to diagrams
with contact terms as shown in fig.~\ref{fig:diag}: the tree level coupling of the
vector mesons to the nucleon via the tensor coupling, diagram (2$^*$), 
enters the magnetic moments as do the LECs $c_6,\,c_7$ in diagram (2), 
diagrams (10) and (10$^*$) as well as (11) and (11$^*$) 
yield vertex corrections to these due to pion loops,
the vector coupling in diagram (3$^*$) compares to the LECs $d_6,\,d_7$ as shown in 
diagram (3), and finally the $\rho$--exchange in diagram (12$^*$) contributes
to the $\pi\pi NN$--coupling $\sim c_4$ in diagram (12).

\medskip \noindent
Explicitly, the analytic results for the form factors including vector mesons
can be gained from those with contact terms only by the following replacements:
\beqa
c_6 &\to& \hat{c}_6 
+ g_{\rho NN} \, \kappa_\rho \, \frac{F_\rho M_\rho}{M_\rho^2-t} ~, \label{c6VM} \\
c_7 &\to& \hat{c}_7 
- \frac{g_{\rho NN} \, \kappa_\rho}{2} \, \frac{F_\rho M_\rho}{M_\rho^2-t}
+ \frac{g_{\omega NN} \, \kappa_\omega}{2} \, \frac{F_\omega M_\omega}{M_\omega^2-t}
+ \frac{g_{\phi NN} \, \kappa_\phi}{2} \, \frac{F_\phi M_\phi}{M_\phi^2-t} ~, \label{c7VM}\\
d_6^r &\to& \hat{d}_6^r
- \frac{g_{\rho NN}}{2} \, \frac{F_\rho}{M_\rho} \, \frac{1}{M_\rho^2-t} ~,\label{d6VM}\\
d_7 &\to& \hat{d}_7
- \frac{g_{\omega NN}}{4} \, \frac{F_\omega}{M_\omega} \, \frac{1}{M_\omega^2-t} 
- \frac{g_{\phi NN}}{4} \, \frac{F_\phi}{M_\phi} \, \frac{1}{M_\phi^2-t} \label{d7VM}~.
\eeqa
Especially, these hold for the loop diagrams with pion loops as vertex corrections
to photon couplings via $c_6,\,c_7$, see diagrams (10), (11) in fig.~\ref{fig:diag} and
(10$^*$), (11$^*$) in fig.~\ref{fig:diagVM}. 
At leading order, resonance saturation for $c_6,\,c_7$ has already been investigated
in \cite{BKMpin}.
The vector meson contributions to the
magnetic moments and the electric radii can be found trivially from 
eqs.~(\ref{kappav})--(\ref{rEs2}) by using eqs.~(\ref{c6VM})--(\ref{d7VM}) at $t=0$.
The vector meson contributions to the magnetic radii, finally, yield a more
complicated replacement law for $e_{54},\,e_{74}$ 
due to the aforementioned loop corrections, which, however, at leading order reads
\beqa
e_{54} &\to& \hat{e}_{54} 
+ \frac{1}{8m_N}\biggl\{ g_{\omega NN} \, \kappa_\omega \, \frac{F_\omega}{M_\omega^3}
+ g_{\phi NN} \, \kappa_\phi \, \frac{F_\phi}{M_\phi^3} \biggr\}
+ {\cal O}(\mu^2) ~, 
\label{e54VM}\\
e_{74}^r &\to& \hat{e}_{74}^r 
+ \frac{1}{4m_N}\, g_{\rho NN} \, \kappa_\rho \, \frac{F_\rho}{M_\rho^3} 
+ {\cal O}(\mu^2) ~. 
\label{e74VM}
\eeqa
Finally, resonance saturation for the LEC $c_4$ has also been analyzed in 
\cite{BKMpin}, where it was found that $c_4$ is completely saturated by $\rho$, $\Delta$,
and (a very small) Roper contribution. Here, we only want to replace the 
$\rho$ contribution by its dynamical analogue, therefore setting
\beq
c_4 \, \to \, \hat{c}_4 + \frac{g_{\rho NN}\,\kappa_\rho}{2m_N}
 \,\frac{G_\rho M_\rho}{M_\rho^2-t} 
\, = \, \hat{c}_4 + \frac{\kappa_\rho}{4m_N} \, \frac{M_\rho^2}{M_\rho^2-t} ~,
\label{c4vm}
\eeq
where, in the second step, a universal $\rho$--hadron coupling 
$g = g_{\rho NN} = g_{\rho\pi\pi} \equiv G_\rho M_\rho/F_\pi^2$
and the KSFR relation $M_\rho^2=2g^2F_\pi^2$ \cite{KSFR}
were assumed.

  
\subsection{Results and discussion}  
\label{sec:Vres}  
Before presenting results for the four form factors, including the effects of dynamical vector 
mesons, we have to discuss the values for the miscellaneous coupling constants introduced  
thereby. First of all, for the vector meson masses we use $M_\rho = 770$ MeV,   
$M_\omega = 780$ MeV, $M_\phi = 1020$ MeV. The couplings to the photon have been fixed from   
the partial widths of $V\rightarrow e^+ e^-$  (in analogy to \cite{mmd})  
to be $F_\rho=152.5$ MeV, $F_\omega=45.7$ MeV, $F_\phi=79.0$ MeV,   
the ratios being in satisfactory agreement with what one would expect from exact   
SU(3)--symmetry (plus ideal $\phi$--$\omega$ mixing),   
$F_V = F_\rho = 3 F_\omega = 3/\sqrt{2} F_\phi$.  
  
\begin{figure}[htb] 
\centerline{ 
\epsfysize=4.2in 
\epsffile{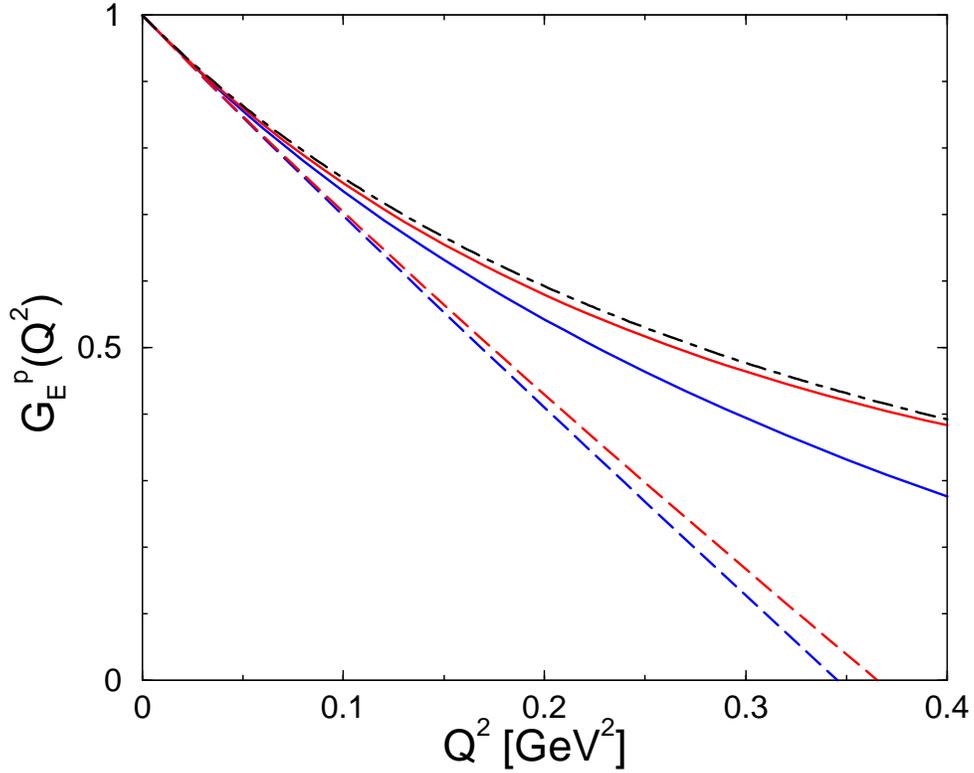} 
} 
\vskip 2.5cm 
\caption{The proton electric form factor in relativistic baryon chiral 
perturbation theory including vector mesons (solid lines) to third (blue curve)  
and fourth (red curve) order. For comparison, the results without vector mesons 
are also shown (blue/red dashed line: third/fourth order). All LECs are 
determined by a fit to the proton charge radius as given by the dispersion 
theoretical result (black dot--dashed curve). 
\label{fig:GEpV} 
} 
\end{figure}

\begin{figure}[htb] 
\centerline{ 
\epsfysize=4.2in 
\epsffile{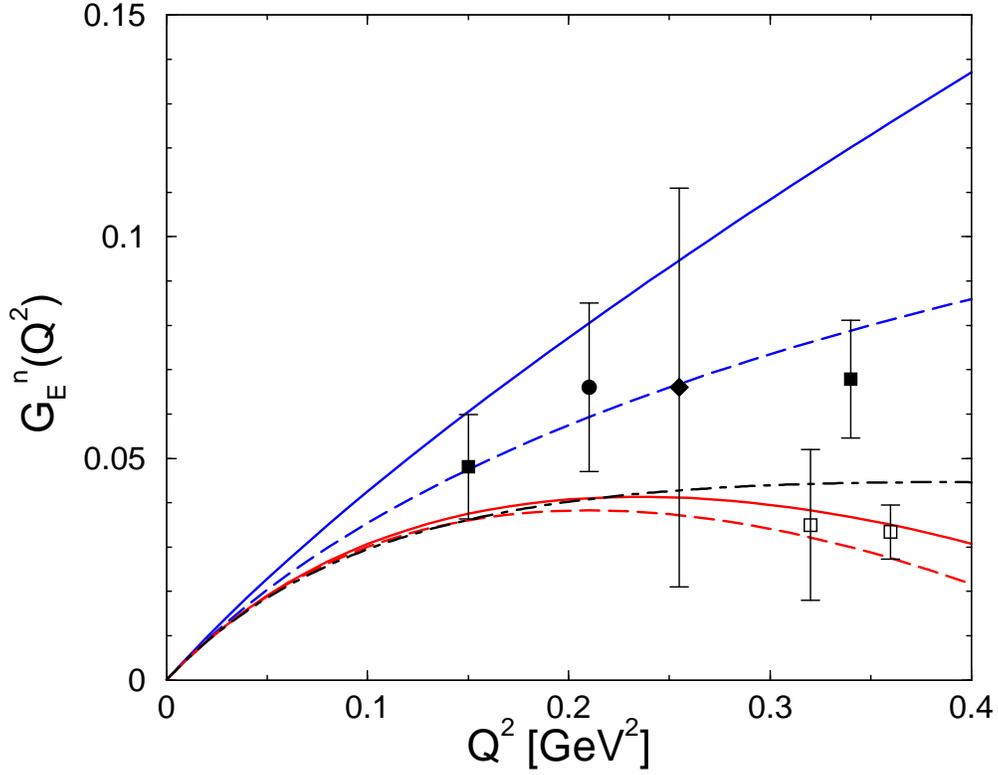} 
} 
\vskip 2.5cm 
\caption{The neutron electric form factor in relativistic baryon chiral 
perturbation theory including vector mesons (solid lines) to third (blue curve)  
and fourth (red curve) order. For comparison, the results without vector mesons 
are also shown (blue/red dashed line: third/fourth order). All LECs are 
determined by a fit to the neutron charge radius. Also given is the result 
of  the dispersion theoretical analysis (black dot--dashed curve). The data 
are from \protect\cite{newnff}. 
\label{fig:GEnV} 
} 
\end{figure}

\begin{figure}[htb] 
\centerline{ 
\epsfysize=4.2in 
\epsffile{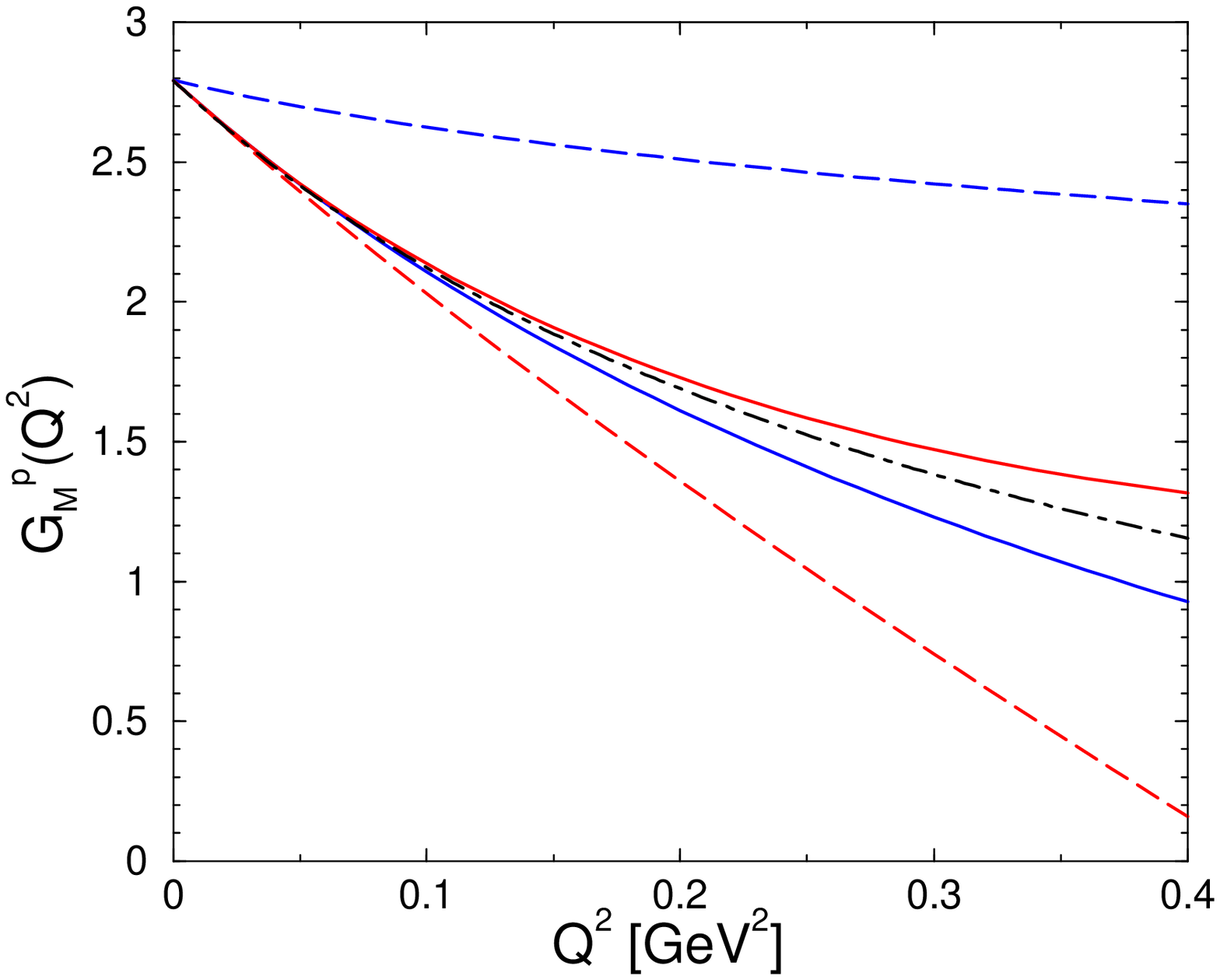} 
} 
\vskip 2.5cm 
\caption{The proton magnetic form factor in relativistic baryon chiral 
perturbation theory including vector mesons (solid lines) to third (blue curve)  
and fourth (red curve) order. For comparison, the results without vector mesons 
are also shown (blue/red dashed line: third/fourth order). All LECs at third (fourth) 
order are determined by a fit to the proton magnetic moment  (radius). 
Also  given is the dispersion theoretical result (black dot--dashed curve). 
\label{fig:GMpV} 
} 
\end{figure}

\begin{figure}[htb] 
\centerline{ 
\epsfysize=4.2in 
\epsffile{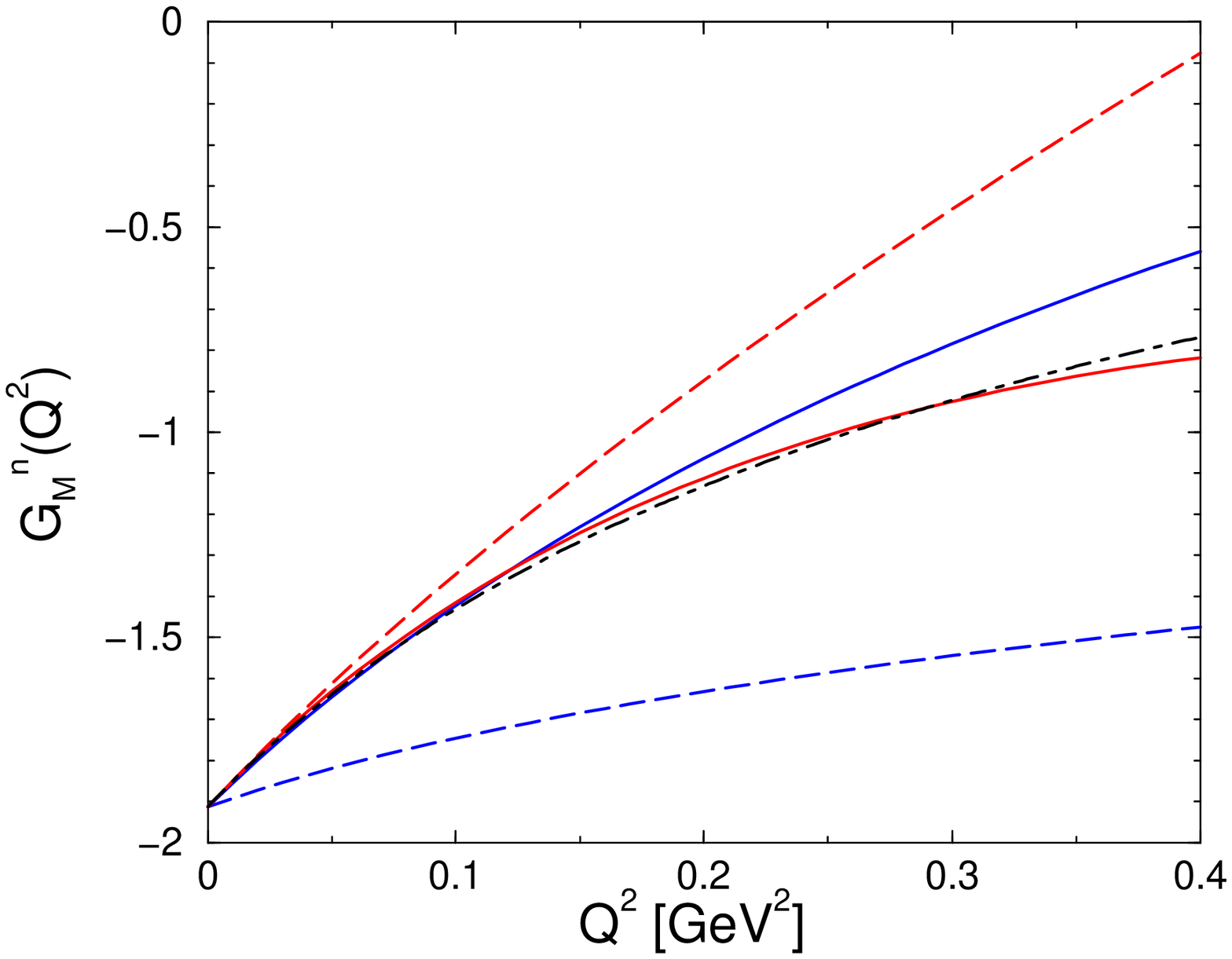} 
} 
\vskip 2.5cm 
\caption{The neutron magnetic form factor in relativistic baryon chiral 
perturbation theory including vector mesons (solid lines) to third (blue curve)  
and fourth (red curve) order. For comparison, the results without vector mesons 
are also shown (blue/red dashed line: third/fourth order). All LECs at third (fourth) 
order are determined by a fit to the neutron magnetic moment  (radius). 
Also  given is  the dispersion theoretical result (black dot--dashed curve). 
\label{fig:GMnV} 
} 
\end{figure}

\medskip \noindent  
The couplings of vector mesons to nucleons are taken from 
the most recent  dispersive analysis \cite{mmd}. 
The respective values (adjusted to our conventions) are
\beq
\begin{array}{lrlr}
g_{\rho NN}  \,=&\!\!  4.0 ~, & \quad \kappa_\rho  \,=&\!\! 6.1~\, ~, \\
g_{\omega NN}\,=&\!\! 41.8 ~, & \quad \kappa_\omega\,=&\!\!-0.16  ~, \\
g_{\phi NN}  \,=&\!\!-18.3 ~, & \quad \kappa_\phi  \,=&\!\!-0.22  ~. 
\end{array}
\eeq
The resulting electric and magnetic form factors of the proton and the
neutron are shown in figs.~\ref{fig:GEpV}--\ref{fig:GMnV}. 
Consider the electric form factors first. 
The vector meson contribution already supplies sufficient curvature 
for an adequate description of the proton charge form factor at third order, 
cf.\ fig.~\ref{fig:GEpV}. This is achieved
without new adjustable parameters, but simply due to the higher order terms induced
by the  inclusion of the vector mesons. It is worth noting that at fourth
order, the chiral plus vector meson representation is in almost perfect agreement
with the result of the dispersive analysis up to momenta of $Q^2 =
0.4\,$GeV$^2$, i.e.\ for much higher momenta than considered so far in chiral
perturbation theory approaches. 
However the already well--described neutron electric form factor, see fig.~\ref{fig:GEnV}, 
is, at fourth order, only very mildly affected by the vector
meson contribution, but is of course also most sensitive to the precise choice
of the vector meson couplings. We only want to state that for a reasonable choice
of these parameters, the good result of the chiral
one--loop representation is not spoiled.\footnote{We also note that the
  dispersive analysis of \cite{mmd} does not include most of the new
  data~\cite{newnff}. Refitting the vector meson parameters accordingly to the
  new data base will also lead to an improvement of the fourth order curve in our analysis.}

\medskip \noindent
Turning now to the magnetic form factors, see figs.~\ref{fig:GMpV},~\ref{fig:GMnV},
we find that both the third and the fourth order curves yield reasonable descriptions
of the data, in both cases the fourth order results being slightly better than the
respective third order ones. The latter show the right slope at the origin (which is
fitted in the fourth order case), and all curves have approximately the right curvature, 
which is a bit stronger at fourth order.

\begin{figure}[htb] 
\centerline{ 
\epsfysize=4.2in 
\epsffile{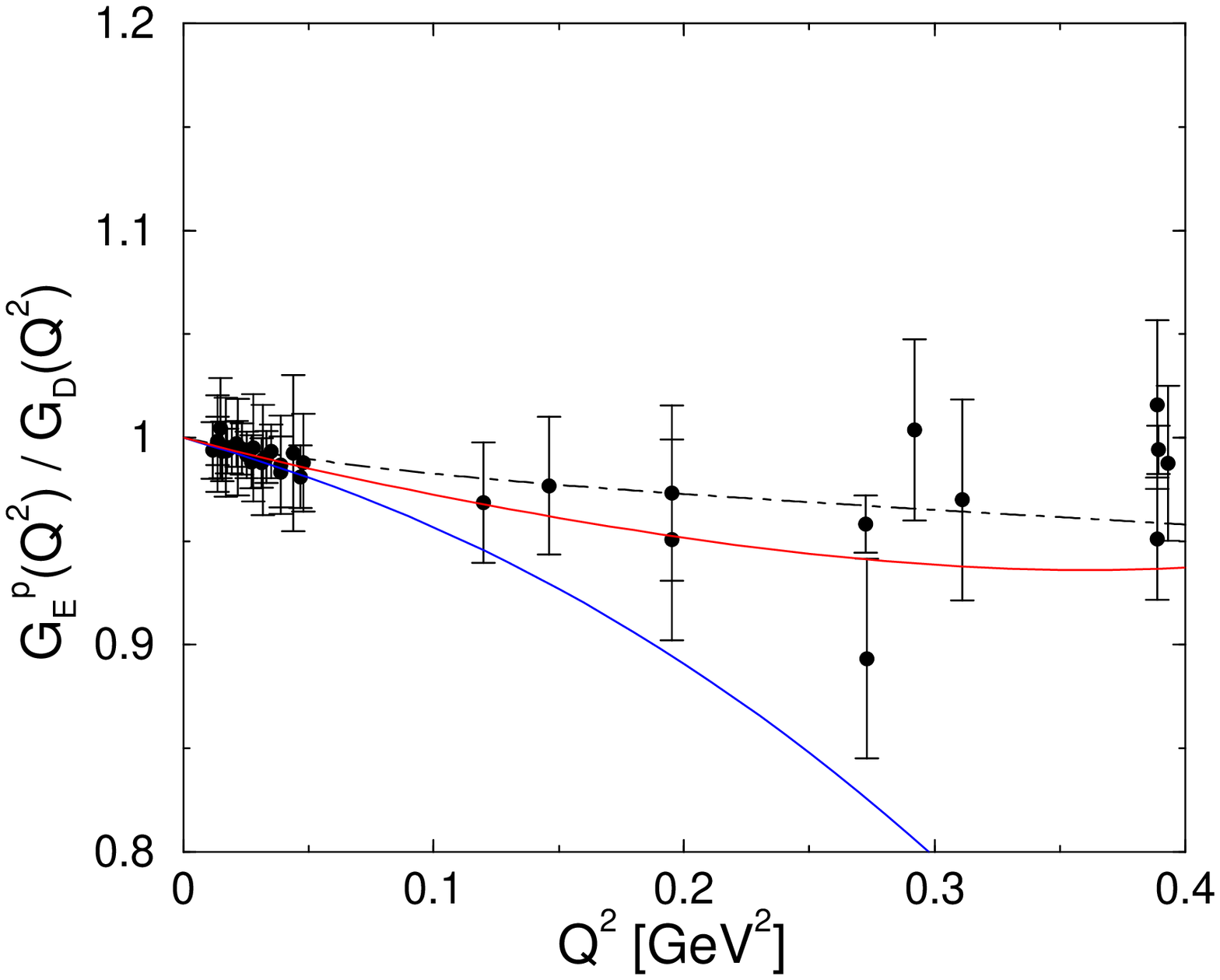} 
} 
\vskip 2.5cm 
\caption{The proton electric form factor in relativistic baryon chiral 
perturbation theory including vector mesons to third (blue curve)  
and fourth (red curve) order, divided by the dipole form factor. 
For comparison, we show the dispersion 
theoretical result (black dot--dashed curve) 
and the world data available in this energy range. 
\label{fig:GEpdip} 
} 
\end{figure}

\begin{figure}[htb] 
\centerline{ 
\epsfysize=4.2in 
\epsffile{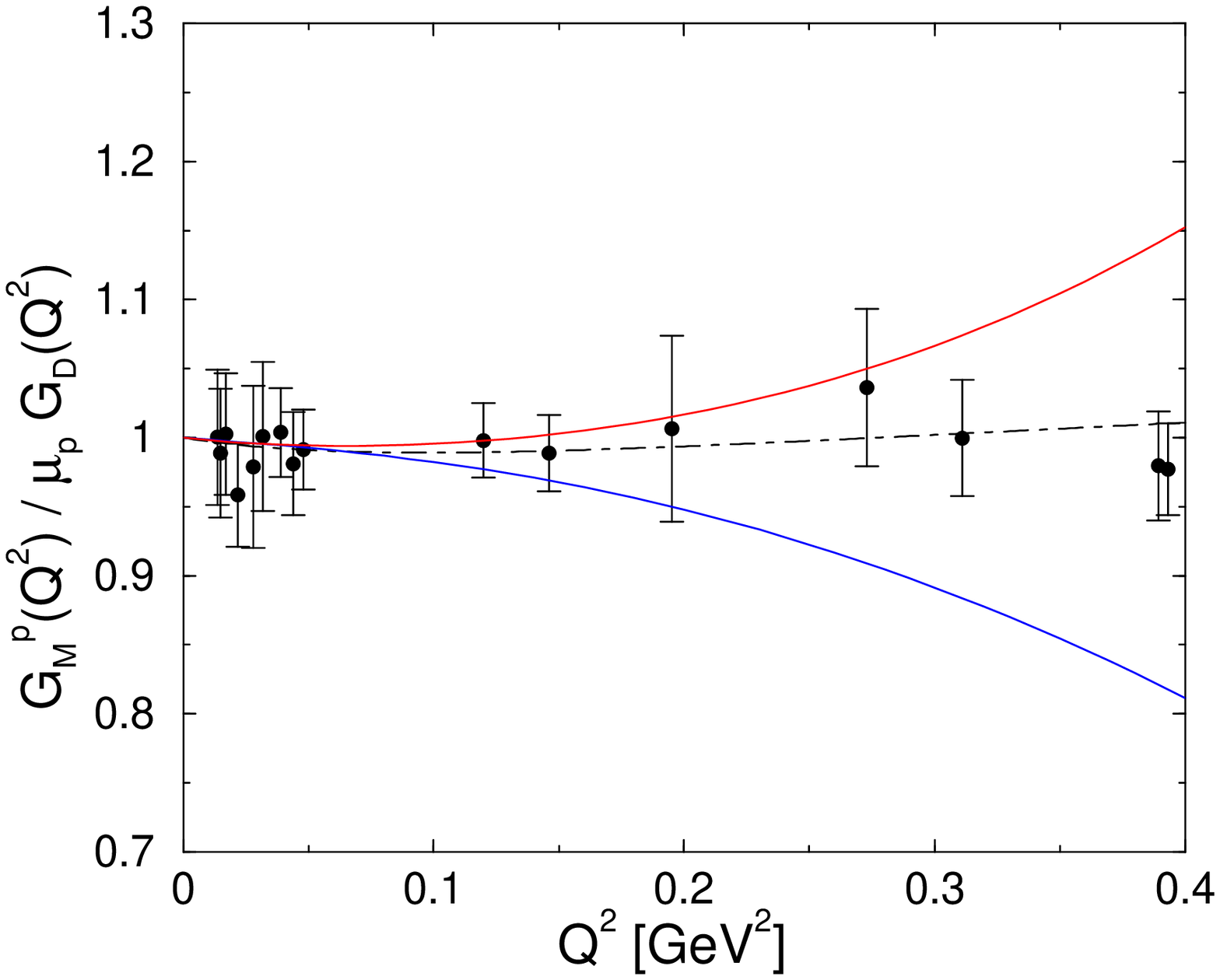} 
} 
\vskip 2.5cm 
\caption{The proton magnetic form factor in relativistic baryon chiral 
perturbation theory including vector mesons to third (blue curve)  
and fourth (red curve) order, divided by the dipole form factor. 
For comparison, we show the dispersion theoretical result (black dot--dashed curve)
and the world data available in this energy range. 
\label{fig:GMpdip} 
} 
\end{figure}

\begin{figure}[htb] 
\centerline{ 
\epsfysize=4.2in 
\epsffile{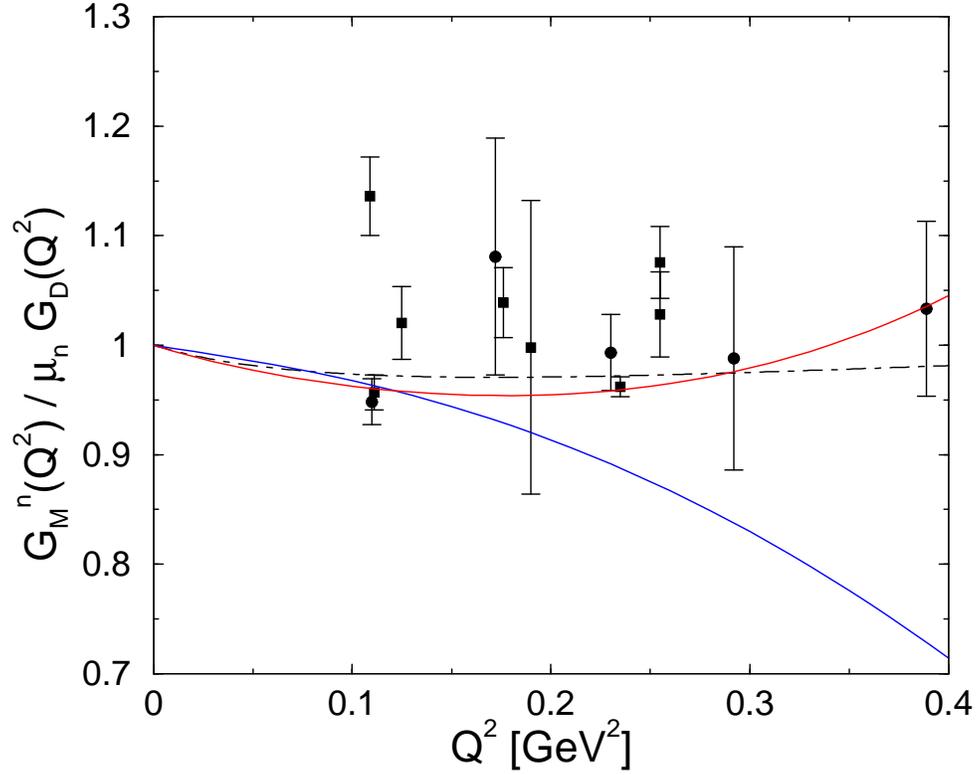} 
} 
\vskip 2.5cm 
\caption{The neutron magnetic form factor in relativistic baryon chiral 
perturbation theory including vector mesons to third (blue curve)  
and fourth (red curve) order, divided by the dipole form factor.
For comparison, we show
the dispersion theoretical result (black dot--dashed curve) 
and the world data available in this energy range, where the data points
denoted by squares (instead of circles) refer to the more recent 
measurements~\protect\cite{gmnnew}. The older data can be traced back
from~\protect\cite{mmd}. 
\label{fig:GMndip} 
} 
\end{figure}

\medskip \noindent
We have, in addition, plotted the third and fourth order curves for $G_E^p$, $G_M^p$,
and $G_M^n$ (including vector mesons), normalized and divided by the dipole form factor
\beq
G_D(Q^2)= \frac{1}{\Bigl(1+Q^2/0.71\,{\rm GeV}^2\Bigr)^2} ~~,
\eeq
see figs.~\ref{fig:GEpdip}--\ref{fig:GMndip}. Here, again, we see that the 
fourth order prediction for the proton charge form factor agrees extremely well 
with the data up to $Q^2=0.4\,$GeV$^2$. 
For the neutron magnetic form factor,  the fourth order curve, though rising above
dipole and dispersion theoretical fit, still is within the error range given by the data,
while the fourth order result for the proton magnetic form factor deviates from
the data above $Q^2=0.3\,$GeV$^2$.

\medskip \noindent
It is worth pointing out the following essential conclusions drawn from this discussion:
\begin{enumerate}
\item In section~\ref{sec:res} we saw that the ``prediction'' of the magnetic radii
in a third order calculation worsens considerably when comparing the relativistic
and the heavy baryon approach. This observation indicated that the large contribution
of the pion cloud to the magnetic radii was an artifact of the $1/m$--expansion 
to ${\cal O}(q^3)$. Inspection of the third order curves including explicit vector meson
effects shows that this reduction of the pion cloud effect is indeed consistent
with the fairly conventional vector meson parameters, yielding in total excellent results
for the magnetic radii which are still ``predicted'' and not fitted. We here show
these ``predictions'' with the separate contributions of pion--cloud and vector mesons,
\beqa
(r_M^p)^2 \!&=&\! (r_M^p)^2_\pi + (r_M^p)^2_{\rm VM}
               =  ( 0.18        + 0.49)\,{\rm fm}^2 = (0.82\,{\rm fm})^2 ~,\\
(r_M^n)^2 \!&=&\! (r_M^n)^2_\pi + (r_M^n)^2_{\rm VM}
               =  ( 0.26        + 0.44)\,{\rm fm}^2 = (0.84\,{\rm fm})^2 ~,
\eeqa
in good agreement with the values from dispersion analyses, $r_M^p = 0.836\,$fm and 
$r_M^n = 0.889\,$fm. Note however that,
although we did not fit any new parameters to the magnetic radii, additional experimental
information enters via the vector meson couplings which in turn have been fitted
to the form factors in the dispersive approach.
\item This analysis indicates that a ``strict'' ChPT calculation to higher--than--fourth order
is probably only moderately sensible. Apart from various two--loop contributions which amount
only to vertex corrections of diagrams already present in the one--loop case, the
major ``new'' contribution is the three--pion continuum entering the isoscalar form factors,
which was already shown to yield a negligibly small correction to the 
$\omega/\phi$--peak in the isoscalar spectral function \cite{spectral}.
Therefore, an analysis to ${\cal O}(q^5)$ or  ${\cal O}(q^6)$ would basically result
in fitting new counterterm parameters in order to reproduce the resonance contributions
considered explicitly here. Apart from that, it is so far not known how to generalize
the infrared regularization scheme to higher loop orders, and
therefore for such a higher order calculation one would have to retrieve to the
non--relativistic formalism. 
\end{enumerate}


\subsection{Remarks on resonance saturation}
\label{sec:ressat}
As we have not \emph{assumed} resonance saturation of the various low energy constants,   
but taken the resonance parameters from elsewhere and refitted the contact terms, we  
may now check how well resonance saturation by the lowest lying vector mesons actually  
works. Table~\ref{tab:LECressat} shows the values for the low energy constants, fitted  
at third and fourth order, with and without vector meson contributions.   
\begin{table}[h]  
\begin{center}  
\renewcommand{\arraystretch}{1.2}  
\begin{tabular}{|c|c|c|c|c|}  
\hline  
& $q^3$ & $q^4$ & $q^3+$VM & $q^4+$VM\\  
\hline  
$c_6$      &    5.03 &    4.77 &    0.20 & $-$0.06 \\  
$c_7$      & $-$2.72 & $-$2.55 & $-$0.27 & $-$0.09 \\  
$d_6^r$    &    0.67 &    0.74 &    1.34 &    1.40 \\      
$d_7$      & $-$0.70 & $-$0.69 & $-$0.04 & $-$0.03 \\  
$e_{54}$   &    ---  &    0.26 &    ---  & $-$0.04 \\  
$e_{74}^r$ &    ---  &    1.65 &    ---  & $-$1.15 \\  
\hline  
\end{tabular}  
\renewcommand{\arraystretch}{1.0}  
\caption{Values for the various low energy constants,   
with and without vector meson contributions.
Note that the third and the fourth columns, strictly speaking, refer to 
the LECs $\hat{\alpha}_i$ instead of $\alpha_i$.
$c_6,\,c_7$ are dimensionless, the $d_i$ are given in GeV$^{-2}$,
the $e_i$ in GeV$^{-3}$.}  
\label{tab:LECressat}  
\end{center}  
\end{table}   
The vector meson contribution to the LEC $c_4$ as given in eq.~(\ref{c4vm}) 
numerically amounts to $1.6\,$GeV$^{-1}$ (see also \cite{BKMpin}) which reduces
the value of $\hat{c}_4$ by about 50\%. This remainder which is due to
(mainly) $\Delta$ and a small Roper contribution is still represented by a contact term.

\medskip \noindent
We conclude that the low energy constants entering the magnetic moments ($c_6,\,c_7$) as  
well as those fitted to the \emph{isoscalar} radii, both electric ($d_7$)   
and magnetic ($e_{54}$), are nearly perfectly saturated by the lowest lying   
vector meson nonet, while the contact terms entering the \emph{isovector} radii   
($d_6,\,e_{74}$) are not at all. This can be interpreted to the effect  
that $\omega$ and $\phi$ are already sufficient to describe the isoscalar channel of  
the vector form factors to good accuracy, while  
higher resonances are mandatory for an adequate description of the isovector form factors.  
This is in agreement with  what is found in dispersive analyses.   
However this does not necessarily invalidate our approach, as these higher resonances  
have much larger masses (the $\rho(1450)$ being the lowest lying of them)   
and therefore will hardly contribute to the \emph{curvature} of the form factors   
(which is the change we are looking for, compared to linear contact terms).

\section{Summary}  
\label{sec:sum}  
\def\theequation{\arabic{section}.\arabic{equation}}  
\setcounter{equation}{0}  

We have studied the electromagnetic form factors of the nucleon
in a manifestly Lorentz invariant form of baryon chiral perturbation
theory to one--loop (fourth) order. As discussed, in this scheme based
on the so--called infrared regularization of loop graphs, one is able
to set up a systematic power counting scheme in harmony with the strictures
from analyticity. The pertinent results of our investigation
can be summarized as follows:
\begin{itemize}  
\item[(1)] To fourth order, the neutron and proton electric form factors each contain
one  low--energy constant which can be fixed from the empirical information
on the corresponding radii. This gives a good description of the neutron charge
form factor up to four--momentum transfer squared of
$Q^2 = 0.4\,$GeV$^2$ and, furthermore, exhibits convergence in that
the corrections when going from third to fourth order are small. 
This is in contrast to the heavy baryon expansion and can be traced back to the proper 
resummation of the recoil terms in the relativistic expansion. For the electric form factor
of the proton, the one--loop representation gives too little curvature and thus deviates
from the data already at $Q^2 \simeq 0.2\,$GeV$^2$, similar to the heavy baryon
description. However, no large fourth order corrections are found below  $Q^2 = 0.4\,$GeV$^2$. 
\item[(2)] To third order, the momentum dependence of the magnetic proton and neutron form
factor is given parameter--free. The $1/m$ corrections present in our approach worsen
the prediction for the magnetic radii based on the leading chiral singularities, 
like e.g.\ in the heavy baryon approach. The leading chiral limit behavior is not a 
good approximation for the Goldstone boson contribution to the magnetic radii.
At fourth order, the magnetic radii can be fixed. Again, there is not
enough curvature in the one--loop representation and one observes large corrections
when going from third to fourth order already at $Q^2 \simeq 0.1\,$GeV$^2$.
\item[(3)] We have  demonstrated explicitly that the spectral
  functions of the isovector form factors have the correct threshold
  behavior. The strong momentum--dependence of these spectral
  functions close to threshold is due to the branch point singularity on the second
  Riemann sheet inherited from the $\pi\pi \to \bar{N}{N}$ P--wave partial
  wave amplitudes.
\item[(4)] We have included the low--lying vector mesons $\rho$, $\omega$, $\phi$ in a
chirally symmetric manner based on an antisymmetric tensor field representation.
This does not introduce any new parameters since these
(masses and coupling constants) are taken from the PDG tables and
from a dispersion theoretical analysis. Refitting the previously defined low--energy
constants by subtracting the vector meson contribution, we find a good description
of {\em all four} form factors already at third order, with small fourth order contributions,
which further improve the theoretical description. In particular, 
we demonstrate that the vector meson
contributions cancel to a large extent in the neutron charge form factor, thus solidifying 
the result obtained in the chiral expansion.
\item[(5)] The inclusion of vector mesons allows to investigate the resonance saturation
hypothesis for these couplings. We find that the couplings related to the magnetic
moments and the isoscalar radii are almost completely saturated by the low--lying vector
mesons. This is, however, not the case for the LECs entering the isovector radii. This
can be traced back to the fact that while the  $\omega$ and the $\phi$ already 
give a good description
of the isoscalar form factors, for the isovector ones one has to include higher mass
states than the $\rho$, in agreement with findings from dispersion theory.
\end{itemize}

\section*{Acknowledgements}
We are grateful to Thomas Becher, V\'eronique Bernard, Nadia Fettes,
Hans--Werner Hammer, and Thomas Hemmert for useful comments and communications.
  


\appendix  
  
\section{Loop integrals}  \label{app:loopfkt}
\def\theequation{\thesection.\arabic{equation}}  
\setcounter{equation}{0}  

In this appendix, we define the loop integrals needed in this paper
and evaluate them in the infrared regularization scheme. Several of
these results have already been given in \cite{Becher}.
  
\subsection{Definition of the loop integrals}  
  
We use the following notation:  
\[  
p'_\mu+p_\mu=Q_\mu ~, \qquad p'_\mu-p_\mu=q_\mu ~, \qquad  
t=q^2~, \]  
\[ \mu = \frac{M_\pi}{m_N}~, \qquad \tau = \frac{t}{m_N^2} ~, \qquad \theta=\frac{t}{M_\pi^2}~.  
\]  
In addition, everywhere except in the loop integrals with just one meson  
and one nucleon propagator, we only need the case where the nucleon momenta are on--shell,  
i.e.\ $p^2=p'^2=m_N^2$.

\medskip \noindent 
Define the following loop integrals:  
\beqa  
\frac{1}{i} \int_I \frac{d^dk}{(2\pi)^d} 
   \frac{1}{M_\pi^2-k^2} &=& \Delta_\pi  ~~,\\  
\frac{1}{i} \int_I \frac{d^dk}{(2\pi)^d}
   \frac{1}{[M_\pi^2-k^2][M_\pi^2-(k+q)^2]} &=& J(t) ~~,\\  
\frac{1}{i} \int_I \frac{d^dk}{(2\pi)^d}
   \frac{k_\mu}{[M_\pi^2-k^2][M_\pi^2-(k+q)^2]} &=&   
  -\frac{1}{2}q_\mu J(t) ~~,\\  
\frac{1}{i} \int_I \frac{d^dk}{(2\pi)^d}
  \frac{k_\mu k_\nu}{[M_\pi^2-k^2][M_\pi^2-(k+q)^2]} &=&   
  \Bigl(q_\mu q_\nu-g_{\mu\nu}t\Bigr)\,J^{(1)}(t) \no\\  
 &+& q_\mu q_\nu \, J^{(2)}(t) ~~,\\  
\frac{1}{i} \int_I \frac{d^dk}{(2\pi)^d}
  \frac{1}{[M_\pi^2-k^2][m_N^2-(p-k)^2]} &=& I(p^2) ~~,\\  
\frac{1}{i} \int_I \frac{d^dk}{(2\pi)^d}
   \frac{k_\mu} {[M_\pi^2-k^2][m_N^2-(p-k)^2]}   
  &=& p_\mu \,I^{(1)}(p^2)  ~~,\\  
\frac{1}{i} \int_I \frac{d^dk}{(2\pi)^d}
  \frac{1}{[M_\pi^2-k^2][m_N^2-(p-k)^2][m_N^2-(p'-k)^2]}   
  &=& I_A(t) ~~,\\  
\frac{1}{i} \int_I \frac{d^dk}{(2\pi)^d}
  \frac{k_\mu}{[M_\pi^2-k^2][m_N^2-(p-k)^2][m_N^2-(p'-k)^2]}   
  &=& Q_\mu \, I_A^{(1)}(t) ~~,\\  
\frac{1}{i} \int_I \frac{d^dk}{(2\pi)^d}
  \frac{k_\mu k_\nu}{[M_\pi^2-k^2][m_N^2-(p-k)^2][m_N^2-(p'-k)^2]}   
  &=& g_{\mu\nu} \, I_A^{(2)}(t) \no\\  
  &+& Q_\mu Q_\nu \, I_A^{(3)}(t)\no\\   
  &+& q_\mu q_\nu \, I_A^{(4)}(t) ~~,\\  
\frac{1}{i} \int_I \frac{d^dk}{(2\pi)^d}
  \frac{1}{[M_\pi^2-k^2][M_\pi^2-(k+q)^2][m_N^2-(p-k)^2]}  
  &=& I_{21}(t) ~~,\\  
\frac{1}{i} \int_I \frac{d^dk}{(2\pi)^d} 
  \frac{k_\mu}{[M_\pi^2-k^2][M_\pi^2-(k+q)^2][m_N^2-(p-k)^2]}  
  &=& Q_\mu\,I_{21}^Q(t) \no\\  
  &-& \frac{1}{2}q_\mu\,I_{21}(t) ~~,\\  
\frac{1}{i} \int_I \frac{d^dk}{(2\pi)^d}
  \frac{k_\mu k_\nu}{[M_\pi^2-k^2][M_\pi^2-(k+q)^2][m_N^2-(p-k)^2]}  
  &=& g_{\mu\nu}\,I_{21}^{00}(t) \no\\  
  &+& Q_\mu Q_\nu\,I_{21}^{QQ}(t) \no\\  
  &+& q_\mu q_\nu \, I_{21}^{qq}(t) \\
  &-& \Bigl(q_\mu Q_\nu+q_\nu Q_\mu \Bigr)\,\frac{1}{2}I_{21}^Q(t)  ~~,\no
\eeqa  
where $\int_I$ symbolizes loop integration according to the
infrared regularization scheme. 
  
\subsection{Reduction of the tensorial loop integrals}  
The reduction of the tensorial loop integrals to the corresponding scalar ones
can be performed in the standard way and leads to the following results:
\beqa  
J^{(1)}(t) &=& \frac{1}{4(d-1)\,t}\Bigl\{ (t-4M_\pi^2)J(t)+2\Delta_\pi \Bigr\} ~~,\\  
J^{(2)}(t) &=& \frac{1}{4}J(t)-\frac{1}{2t}\Delta_\pi ~~,\\  
I^{(1)}(p^2) &=& \frac{1}{2p^2}\Bigl\{(p^2-m_N^2+M_\pi^2)I(p^2) + \Delta_\pi \Bigr\} ~~,\\  
I_A^{(1)}(t) &=& \frac{1}{4m_N^2-t}\Bigl\{I(m_N^2)+M_\pi^2I_A(t) \Bigr\} ~~,\\  
I_A^{(2)}(t) &=& \frac{1}{d-2}\Bigl\{I_A(t)  
  -I_A^{(1)}(t)\Bigr\}M_\pi^2 ~~,\\  
I_A^{(3)}(t) &=&\frac{1}{(d-2)(4m_N^2-t)}\biggl\{\Bigl((d-1)  
  I_A^{(1)}(t)-I_A(t)\Bigr)M_\pi^2+\frac{d-2}{2}I^{(1)}(m_N^2) \biggr\} ~,~\\  
I_A^{(4)}(t) &=&\frac{1}{(d-2)\,t}\biggl\{ \Bigl(I_A^{(1)}(t)  
  -I_A(t)\Bigr)M_\pi^2-\frac{d-2}{2}I^{(1)}(m_N^2)  \biggr\} ~~,\\  
I_{21}^Q(t) &=& \frac{1}{2(4m_N^2-t)}\Bigl\{(2M_\pi^2-t)I_{21}(t)  
  -2I(m_N^2)+2J(t)\Bigr\} ~~,\\  
I_{21}^{00}(t) &=& \frac{1}{4(2-d)}\Bigl\{ 2I(m_N^2)-(4M_\pi^2-t)I_{21}(t)  
  +2(2M_\pi^2-t)I_{21}^Q(t) \Bigr\} ~~,\\  
I_{21}^{QQ}(t) &=& \frac{1}{4(d-2)(4m_N^2-t)}\biggl\{   
  2I(m_N^2)-2(d-2)I^{(1)}(m_N^2) \no\\ && \qquad\qquad  
  -(4M_\pi^2-t)I_{21}(t)+2(d-1)(2M_\pi^2-t)I_{21}^Q(t) \biggr\} ~~,\\  
I_{21}^{qq}(t) &=& \frac{1}{4(d-2)t}\biggl\{  
  -2(d-3)I(m_N^2)+2(d-2)I^{(1)}(m_N^2) \no\\ && \qquad\qquad  
  -\Bigl(4M_\pi^2-(d-1)\,t\Bigr)I_{21}(t)+2(2M_\pi^2-t)I_{21}^Q(t) \biggr\}  ~~.
\eeqa  
  
\subsection{Scalar loop integrals}  
The scalar loop integrals are found to be  
\beqa  
\Delta_\pi &=& 2M_\pi^2\biggl\{ L+\frac{1}{16\pi^2}\log\mu\biggr\} ~~,\\  
J(t) &=& -2 \biggl\{ L+\frac{1}{16\pi^2}\log\mu\biggr\}  
  - \frac{1}{16\pi^2} \Bigl(1+k(t)\Bigr) ~~,\\  
I(m_N^2) &=& -\mu^2 \biggl(L+\frac{1}{16\pi^2}\log\mu\biggr)  
  +\frac{\mu}{16\pi^2}\biggl\{ \frac{\mu}{2}-\sqrt{4-\mu^2}  
  \arccos\biggl(-\frac{\mu}{2}\biggr) \biggr\} ~~,\\   
I_A(t) &=& -\frac{f(t)}{m_N^2}\biggl\{ L+\frac{1}{16\pi^2}  
  \biggl(\log\mu+\frac{1}{2}\biggr)\biggr\}  
  +\frac{1}{16\pi^2}\frac{\mu}{2m_N^2}g(t) ~~,\\  
I_{21}(t) &=& \frac{f(t)}{m_N^2}\biggl\{L+\frac{1}{16\pi^2}  
  \biggl(\log\mu+\frac{1}{2}\biggr)\biggr\}  
  +\frac{1}{16\pi^2}\frac{1}{2m_N^2}\biggl(h_1(t)+\frac{2-\mu^2}{\mu}h_2(t)\biggr)  ~,~
\eeqa  
where the following loop functions have been reduced to integrals over  
one Feynman parameter:  
\beqa  
k(t)&=&\int_0^1 dx \log\Bigl(1-x(1-x)\,\theta\Bigr) 
= \sqrt{\frac{4-\theta}{-\theta}} \log \biggl( 
  \frac{\sqrt{4-\theta}+\sqrt{-\theta}}{\sqrt{4-\theta}-\sqrt{-\theta}} \biggr)
  -2 ~~,\\  
&&\no\\  
f(t)&=&\int_0^1 dx \frac{dx}{1-x(1-x)\,\tau}  
  \,=\, \frac{2}{\sqrt{-\tau\smash{(4-\tau)}}}   
\log\biggl(\frac{\sqrt{4-\tau}+\sqrt{-\tau}}{\sqrt{4-\tau}-\sqrt{-\tau}}\biggr) ~~,\\  
&&\no\\  
g(t)&=& \int_0^1 dx \frac{ \arccos \Bigl( - \frac{\mu}{2\sqrt{1-x(1-x)\tau}}   
  \Bigr) }  
 {\Bigl(1-x(1-x)\,\tau\Bigr)\sqrt{1-\frac{\mu^2}{4}-x(1-x)\,\tau} } ~~,\\  
&&\no\\  
h_1(t) &=& \int_0^1 dx \frac{\log\Bigl(1-x(1-x)\,\theta\Bigr)}  
  {1-x(1-x)\,\tau} ~~,\\  
&&\no\\  
h_2(t)&=& \int_0^1 dx \frac{ \arccos \Bigl( -   
  \frac{\mu\bigl(\frac{1}{2}-x(1-x)\,\theta\bigr)}  
       {\sqrt{(1-x(1-x)\,\tau)\,(1-x(1-x)\,\theta)}}   
 \Bigr) }  
 {\Bigl(1-x(1-x)\,\tau\Bigr)\sqrt{1-\frac{\mu^2}{4}-x(1-x)\,\theta} }  ~~.
\eeqa  

  
\section{Form factor contributions from separate diagrams}  
\def\theequation{\thesection.\arabic{equation}}  
\setcounter{equation}{0}  
\label{app:contri}  
In this section, we give the contributions to the form factors $F_1(t),\,F_2(t)$,
coming from the various diagrams shown in fig.~\ref{fig:diag} .


\subsection{Contributions to $F_1$}  
\beqa  
E_{1+3} &=& 1 - \Bigl( \tau^3\,d_6 + 2\,d_7 \Bigr)\, t ~~,\\
E_5 &=& \frac{g_A^2}{8F_\pi^2} \, (3-\tau^3) \, \Bigl\{  
 \Delta_\pi - 4m_N^2 I^{(1)}(m_N^2) - 4m_N^2M_\pi^2I_A(t) \no\\  
&& \qquad\qquad\qquad + 8m_N^2 I_A^{(2)}(t)   
 +32m_N^4 I_A^{(3)}(t) \Bigr\} ~~,\\  
E_6 &=& -\frac{g_A^2}{F_\pi^2} \, \tau^3 \, \Bigl\{ t J^{(1)}(t)   
 +4m_N^2 I_{21}^{00}(t) +16m_N^4 I_{21}^{QQ}(t) \Bigr\} ~~,\\  
E_7 &=& \frac{g_A^2}{F_\pi^2}\,\tau^3\,\Bigl\{\Delta_\pi-2m_N^2I^{(1)}(m_N^2) \Bigr\} ~~,\\  
E_8 &=& -\frac{\tau^3}{2F_\pi^2}\,\Delta_\pi ~~,\\  
E_9 &=& \frac{\tau^3}{F_\pi^2}\,t \,J^{(1)}(t) ~~,\\  
E_{10} &=& \frac{m_N^2g_A^2}{F_\pi^2}\Bigl( (3-\tau^3)c_6 +6c_7\Bigr)\,t\,I_A^{(3)}(t) ~~,\\  
E_{11} &=& \frac{3M_\pi^2}{4m_NF_\pi^2}\,(1+\tau^3)\,c_2\,\biggl\{\Delta_\pi   
  - \frac{M_\pi^2}{32\pi^2} \biggr\}  ~~.
\eeqa  
  

\subsection{Contributions to $F_2$}  
\beqa
M_{2+3+4} &=& \frac{1}{2}(1+\tau^3)\,c_6 + c_7
  + \Bigl( \tau^3\,d_6 + 2\,d_7 \Bigr)\, t
  + 2m_N \Bigl( 2\,e_{54} + \tau^3\,e_{74}  \Bigr)\, t \no\\
&& - 8m_N M_\pi^2 \Bigl( 2\,e_{105}+\tau^3\,e_{106} \Bigr) ~~,\\
M_5 &=& -\frac{g_A^2}{F_\pi^2} \, (3-\tau^3) \, 4 m_N^4 \, I_A^{(3)}(t) ~~,\\  
M_6 &=& \frac{g_A^2}{F_\pi^2} \, \tau^3 \, 16 m_N^4I_{21}^{QQ}(t) ~~,\\  
M_{10} &=& -\frac{m_N^4g_A^2}{8F_\pi^2}\Bigl( (3-\tau^3)\,c_6 +6c_7\Bigr)  
 \Bigl\{ \Delta_\pi -4m_N^2 I^{(1)}(m_N^2) +4m_N^2M_\pi^2 I_A(t) \no\\  
&& \qquad -16m_N^2I_A^{(2)}(t) +8m_N^2t\Bigl( I_A^{(3)}(t)-I_A^{(4)}(t) \Bigr)  
 \Bigr\} ~~,\\  
M_{11} &=&  -\frac{3M_\pi^2}{4m_NF_\pi^2}\,(1+\tau^3)\,c_2\,\biggl\{\Delta_\pi   
  - \frac{M_\pi^2}{32\pi^2} \biggr\}  
  -\frac{\tau^3}{2F_\pi^2} \,c_6\, \Delta_\pi ~~,\\  
M_{12} &=& \frac{4m_N}{F_\pi^2} \,\tau^3 \, c_4 \, t\, J^{(1)}(t)  ~~.
\eeqa  
  

\subsection{Z--factor}  
We here spell out the Z--factor needed for wave function renormalization up to
fourth order. For the evaluation of the form factor $F_1$, the nucleon charge
has to be multiplied by $Z_N$, while for $F_2$, the anomalous magnetic
moment (a second order contribution) only has to be renormalized by 
the Z--factor up to third order, i.e.\ the term $\sim c_2$ can be dropped.
\beqa  
Z_N &=& 1 \,+ \, \frac{3g_A^2}{4F_\pi^2} \biggl\{ 
M_\pi^2 I(m_N^2) - 2m_N^2 I^{(1)}(m_N^2) 
   + 4 m_N^2 M_\pi^2 \Bigl( I_A(0) -2I_A^{(1)}(0)\Bigr) \biggr\}\no\\  
&& \quad - \, c_2 \, \frac{3M_\pi^2}{2m_NF_\pi^2}
   \biggl(\Delta_\pi - \frac{M_\pi^2}{32\pi^2} \biggr)  ~~.
\eeqa  


\section{Imaginary parts and spectral functions}
\def\theequation{\thesection.\arabic{equation}}  
\setcounter{equation}{0}  
\label{app:spectral}  

\begin{figure}[htb] 
\centerline{ 
\epsfysize=4.9in 
\epsffile{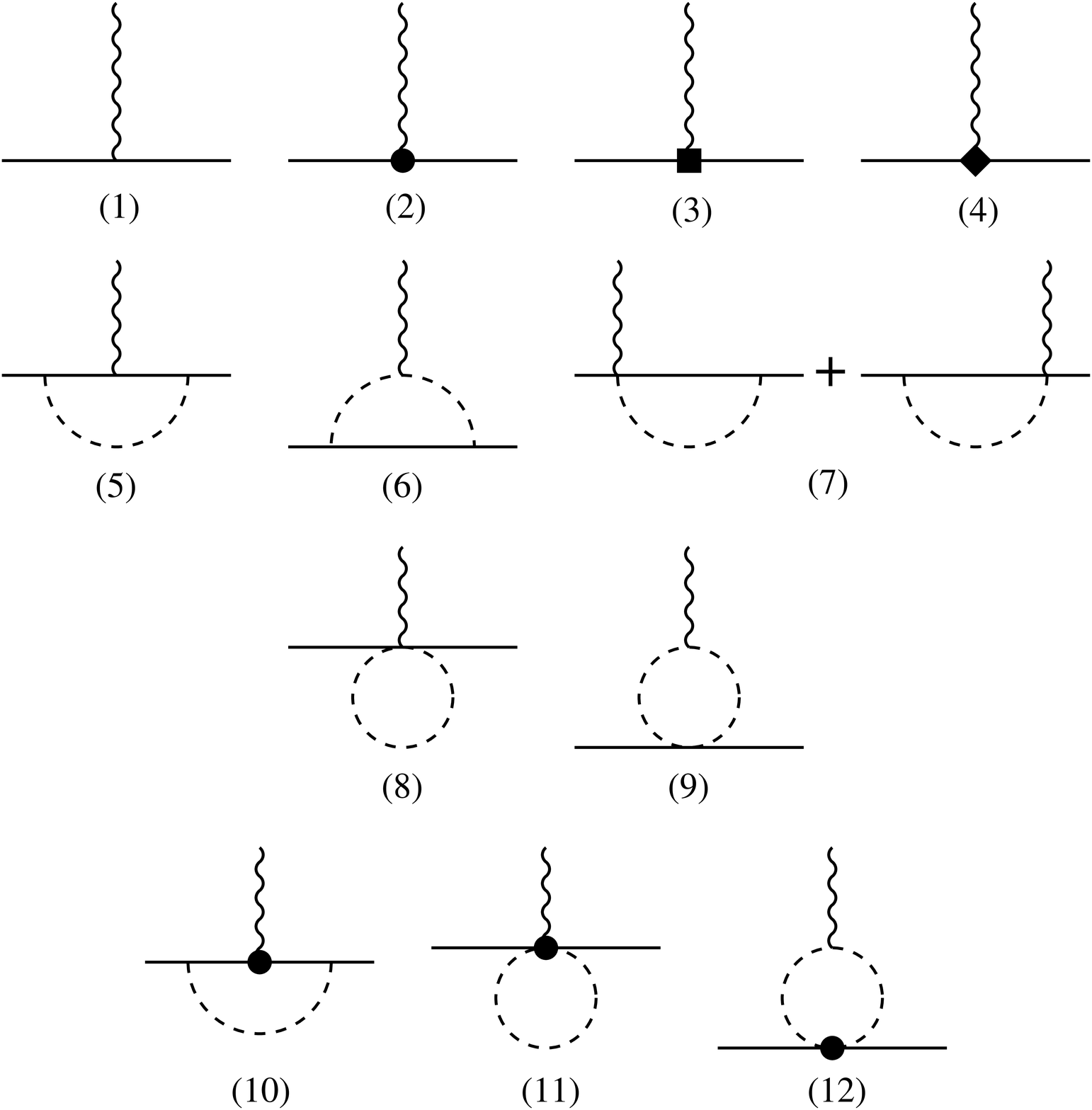} 
}
\vskip 1.5cm 
\caption{Feynman diagrams contributing to the electromagnetic form factors
up to fourth order. Solid, dashed, and wiggly lines refer to nucleons,
pions, and the vector source, respectively. Vertices denoted by a heavy
dot / a square / a diamond refer to insertions from the 
second / third / fourth order chiral Lagrangian, respectively.
Diagrams contributing via wave function renormalization only are not shown.
\label{fig:diag} 
} 
\end{figure}

Out of the diagrams depicted in fig.~\ref{fig:diag}, only graphs (6), (9), 
and (12) contribute to the spectral functions of the isovector form factors
in the low energy region, i.e.\ starting at the threshold $t=4M_\pi^2$. The separate
contributions can be calculated either from the imaginary parts of
the two basic loop functions involved (see also \cite{GSS}), 
\beqa
{\rm Im} \, J(t) &=& \frac{1}{16\pi} \,\sqrt{1-\frac{4M_\pi^2}{t}} ~~, \\
{\rm Im} \, I_{21}(t) &=& \frac{1}{8\pi} \,\frac{1}{\sqrt{t(4m_N^2-t)}} \,
         \arctan \frac{\sqrt{(4m_N^2-t)(t-4M_\pi^2)}}{t-2M_\pi^2} ~~,
\eeqa
or directly by using Cutkosky rules. This leads to the following expressions:
\beqa
{\rm Im} \, E_6 &=& 
  \frac{\tau^3 \, g_A^2}{192\pi F_\pi^2\,(4m_N^2-t)^2}\,\sqrt{1-\frac{4M_\pi^2}{t}} \no\\
&\times& \Biggl\{ 16 m_N^4 \Bigl( 5t-8M_\pi^2 \Bigr) 
  +4m_N^2 \,t\, \Bigl( 5t-14M_\pi^2 \Bigr)
  -t^2 \Bigl(t-4M_\pi^2 \Bigr) \\
&&\quad-48m_N^2 \, \frac{m_N^2\Bigl(3t^2-12M_\pi^2t+8M_\pi^4\Bigr)+M_\pi^4t}
  {\sqrt{(4m_N^2-t)(t-4M_\pi^2)}} \arctan 
  \frac{\sqrt{(4m_N^2-t)(t-4M_\pi^2)}}{t-2M_\pi^2} \Biggr\} \no
  ~~,\\
{\rm Im} \, M_6 &=& 
\frac{\tau^3 \, g_A^2\,m_N^4}{4\pi F_\pi^2\,(4m_N^2-t)^2}\,\sqrt{1-\frac{4M_\pi^2}{t}} \,
  \Biggl\{ -3 \Bigl( t-2M_\pi^2 \Bigr) \no\\
&&\quad+2 \, \frac{\Bigl(2m_N^2 +t \Bigr) \Bigl(t-4M_\pi^2\Bigr)+6M_\pi^4}
  {\sqrt{(4m_N^2-t)(t-4M_\pi^2)}} \arctan 
  \frac{\sqrt{(4m_N^2-t)(t-4M_\pi^2)}}{t-2M_\pi^2} \Biggr\} 
  ~~,\\
{\rm Im} \, E_9 &=& 
  \frac{\tau^3}{192\pi F_\pi^2}\,\frac{(t-4M_\pi^2)^{3/2}}{t^{1/2}} ~~, \\
{\rm Im} \, M_{12} &=&
  \frac{m_N\,c_4\,\tau^3}{48\pi F_\pi^2}\,\frac{(t-4M_\pi^2)^{3/2}}{t^{1/2}} ~~. 
\eeqa
For these results, see also \cite{GSS,spectral}.


\bigskip 


$\,$ 
\vskip 1.5cm 

\noindent {\Large {\bf Figures}} 

\vskip 1.5cm 
 
\pagebreak 
 
$\,$ 
\vskip 2.0cm 
 
\pagebreak
 
$\,$ 
\vskip 3cm 

\pagebreak 
 
$\,$ 
\vskip 3cm

\pagebreak 
 
$\,$ 
\vskip 3cm 
 
\pagebreak 
 
$\,$ 
\vskip 3cm

\pagebreak

$\,$ 
\vskip 4cm 

\pagebreak
 
$\,$ 
\vskip 3cm 
 
\pagebreak 
 
$\,$ 
\vskip 3cm

\pagebreak 
 
$\,$ 
\vskip 3cm 

\pagebreak 
 
$\,$ 
\vskip 3cm

\pagebreak 
 
$\,$ 
\vskip 3cm 
 
\pagebreak 
 
$\,$ 
\vskip 3cm

\pagebreak 
 
$\,$ 
\vskip 3cm

\pagebreak

\end{document}